%% file: main.tex
\documentclass{ar-1col}
\usepackage[numbers]{natbib}
\usepackage{url}
\usepackage{amsmath,amssymb,amsthm}
\usepackage{mathtools}
\usepackage{xspace}
\input{macros-dave}
\usepackage[colorlinks=true]{hyperref}
\jname{Xxxx. Xxx. Xxx. Xxx.}
\jvol{AA}
\jyear{YYYY}
\doi{10.1146/((please add article doi))}

\begin{document}

\markboth{Kwiatkowska et al.}{Probabilistic Model Checking and Autonomy}

\title{Probabilistic Model Checking and Autonomy}

\author{Marta Kwiatkowska,$^1$ Gethin Norman$^2$ and David Parker$^3$
\affil{$^1$Department of Computer Science, University of Oxford, UK, OX1 3QD; email: marta.kwiatkowska@cs.ox.ac.uk}
\affil{$^2$School of Computing Science, University of Glasgow, UK, G12 8RZ; email: gethin.norman@glasgow.ac.uk}
\affil{$^3$School of Computer Science, University of Birmingham, UK; email: d.a.parker@cs.bham.ac.uk}}

\begin{abstract}
Design and control of autonomous systems that operate in uncertain or adversarial environments can be facilitated by formal modelling and analysis. Probabilistic model checking is a technique to automatically verify, for a given temporal logic specification, that a system model satisfies the specification, as well as to synthesise an optimal strategy for its control. This method has recently been extended to multi-agent systems that exhibit competitive or cooperative behaviour modelled via stochastic games and synthesis of equilibria strategies. In this paper, we provide an overview of probabilistic model checking, focusing on models supported by the PRISM and PRISM-games model checkers. This includes fully observable and partially observable Markov decision processes, as well as turn-based and concurrent stochastic games, together with associated probabilistic temporal logics. We demonstrate the applicability of the framework through illustrative examples from autonomous systems. Finally, we highlight research challenges and suggest directions for future work in this area.
\end{abstract}

\maketitle


\input{introduction}
\input{mdps}

\input{pomdps}
\input{tsgs}

\input{csgs}
\input{extensions}

\input{conclusions}

\bibliographystyle{ar-style3}
\bibliography{bib}

\end{document}

%% file: macros-dave.tex
\renewcommand{\emptyset}{\varnothing}

\DeclareMathOperator{\opt}{opt}

\usepackage{scalerel}

\newcommand\scale[2]{\vstretch{#1}{\hstretch{#1}{#2}}}

\newcommand{\sectref}[1]{Section~\ref{#1}}

\newcommand{\figref}[1]{{\bf Figure~\ref{#1}}}

\newcommand{\egref}[1]{Example~\ref{#1}}

\newcommand{\eqnref}[1]{Equation~\ref{#1}}

\newcommand{\defref}[1]{Definition~\ref{#1}}

\newcommand{\sectsectref}[2]{Sections~\ref{#1} and~\ref{#2}}

\newcommand{\defdefref}[2]{Definitions~\ref{#1} and~\ref{#2}}

\theoremstyle{definition} 
\newtheorem{definition}{Definition}
\newtheorem{example}{Example}

\newcounter{exampcount}
{\hfill$\blacksquare$\vskip6pt}

\def\ra{{\rightarrow}}

\def\cC{{\mathcal{C}}}

\def\Nset{\mathbb{N}}

\def\Rset{\mathbb{R}}

\def\Eset{\mathbb{E}}
\def\Rsetgeq{\mathbb{R}_{\geq 0}}

\def\ra{\rightarrow} 
\def\rmdef{\,\stackrel{\mbox{\rm {\tiny def}}}{=}}
\newcommand{\true}{\mathtt{true}} 
\def\E{\mathbb{E}}
\renewcommand{\leq}{\leqslant}
\renewcommand{\geq}{\geqslant}
\renewcommand{\le}{\leqslant}

\newcommand\mdp{{\sf M}}
\newcommand\pomdp{{\sf P}}

\newcommand\tsg{{\sf T}}
\newcommand\csg{{\sf C}}

\newcommand\sinit{{\bar{s}}}

\newcommand\act{\mathit{A}}

\newcommand\acts{\act}
\newcommand\dist{{\mathit{Dist}}}
\newcommand\Dist{{\dist}}

\newcommand\Prob{{\mathit{Prob}}}

\newcommand\strat{{\sigma}}
\newcommand\strats{{\Sigma}}
\newcommand\val{{\mathit{val}}}

\newcommand{\estrat}[2]{\E_{#1}^{#2}} 
\newcommand{\srew}{\rew_{S}} 
\newcommand{\arew}{\rew_{\act}} 

\newcommand{\mdptuple}{{(S,\sinit,A,\delta,\AP,L)}}
\newcommand{\pomdptuple}{{(S,\sinit,A,\delta,\AP,L,\obs,\obsf)}}
\newcommand{\tsgtuple}{{(N,S,(S_i)_{i=1}^n,\sinit,A,\delta,\AP,\lab)}}
\newcommand{\csgtuple}{{(N, S, \sinit, A, \Delta, \delta, \AP, \lab)}}

\newcommand{\last}{\mathit{last}}
\newcommand{\ipaths}{\mathit{IPaths}}
\newcommand{\fpaths}{\mathit{FPaths}}

\newcommand{\fpat}{\pi}

\newcommand{\enab}[1]{A(#1)}

\newcommand{\obs}{\mathcal{O}}
\newcommand{\obsf}{\mathit{obs}}

\def\AP{{\mathit{AP}}}
\def\sat{{\,\models\,}}

\def\sateps{{\,\models\,}}

\newcommand{\coalition}[1]{\langle \! \langle {#1} \rangle \! \rangle}

\def\next{{X\,}}
\def\until{{\ {\cal U}\ }}
\def\buntil{{\ {\cal U}^{\leq k}\ }}

\def\next{{\mathtt X\,}}
\def\until{{\ \mathtt{U}\ }}
\def\untilop{{\mathtt{U}}}

\def\buntil{{\ \mathtt{U}^{\leq k}\ }}

\def\future{{\mathtt{F}\ }}
\def\futureop{{\mathtt{F}}}

\def\bfuture{{\mathtt{F}^{\leq k}\ }}

\def\globally{{\mathtt{G}\ }}
\newcommand\bgloballyp[1]{{\mathtt{G}^{\leq #1}\ }}
\newcommand{\cumul}[1]{\mathtt{C}^{#1}} 

\newcommand{\ap}{\mathsf{a}}

\newcommand{\instant}[1]{\mathtt{I}^{#1}} 
\newcommand{\reachrew}[1]{\mathtt{F}\ {#1}} 

\newcommand\probopP{{\mathtt P}}
\newcommand\nashop[3]{\coalition{#1}_{#2}(#3)}

\newcommand\probop[2]{\probopP_{#1}[\,{#2}\,]}
\newcommand\probopbp[1]{\probop{\bowtie p}{#1}}
\newcommand\rewopR{{\mathtt R}}
\newcommand\rewop[3]{\rewopR^{#1}_{\,#2}[\,{#3}\,]}
\newcommand\rewopbr[2]{\rewop{#1}{\bowtie q}{#2}}

\newcommand{\lab}{{\mathit{L}}}

\newcommand{\rew}{r}

\newcommand{\pctlstar}[1]{PCTL* }
\newcommand{\rpatlstar}[1]{RPATL* }

\newcommand{\coal}[1]{\lla #1 \rra}
\newcommand{\lla}{\langle\!\langle}
\newcommand{\rra}{\rangle\!\rangle}


%% file: introduction.tex
\section{INTRODUCTION}\label{intro-sect}

As autonomous systems become embedded within computing infrastructure, from information systems through to security and robotics, there is a growing need for methodologies that ensure their safe, secure, reliable, timely and resource efficient execution. Design of computer systems can be facilitated by formal modelling and verification, and in particular \emph{model checking}, which aims to automatically check if a system model satisfies given requirements typically expressed in temporal logic. \emph{Autonomy}, however, creates additional demands of controllability, since autonomous systems operate in uncertain or adversarial environments, and strategic reasoning, to ensure effective coordination of cooperative or competitive behaviour of system components (agents). 

\emph{Probabilistic model checking} is a collection of techniques for the modelling of systems that exhibit probabilistic and non-deterministic behaviour, which supports not only their model checking against temporal logic, but also \emph{synthesis of optimal controllers (strategies)} from temporal logic specifications. Probability is used to quantify environmental uncertainty and stochasticity, while non-determinism represents model decisions. Markov decision processes (MDPs) are typically employed to model and reason about the strategic behaviour of an agent against a stochastic environment, where specifications are expressed in probabilistic extensions of the temporal logics CTL or LTL.
Partially observable Markov decision processes (POMDPs) permit similar modelling and analysis, but for contexts where the agent has limited power to observe its environment.

MDPs and POMDPs, however, are unable to faithfully represent the behaviour of multiple players competing or cooperating to achieve their individual goals. To this end, we employ multi-agent systems modelled via \emph{stochastic games} and reason about their \emph{strategic behaviour} for both zero-sum and nonzero-sum (equilibria) properties. For zero-sum properties, the utilities of an agent are the negation of the utility of its opponent, whereas for nonzero-sum each agent is pursuing its own quantitative objective. Probabilistic model checking has been recently extended to encompass both turn-based and concurrent stochastic games, together with an extension of the temporal logic that inherits the coalition operator from ATL, as well as synthesis of optimal \emph{Nash equilibria} strategies (more precisely, subgame-perfect social welfare optimal strategies).

In this paper, we provide an overview of recent advances in probabilistic model checking, focusing on the model checking and strategic reasoning methods implemented in the PRISM \cite{KNP11} and PRISM-games~\cite{KNPS20} tools for discrete probabilistic models. The review covers fully observable and partially observable Markov decision processes (\sectsectref{mdps-sect}{pomdps-sect} respectively), as well as turn-based and concurrent stochastic games (\sectsectref{tsgs-sect}{csgs-sect} respectively), together with associated probabilistic temporal logics. We discuss the core types of quantitative analyses available for each model, as well as extensions such as multi-objective analysis and continuous-time, also called real-time in the model checking literature, models (\sectref{Extensions}). We demonstrate the applicability of the framework through illustrative examples, with emphasis on the areas of robotics and autonomy. Finally, we highlight challenges and suggest directions for future work in this area (\sectref{conc-sect}).

%% file: mdps.tex
\section{MARKOV DECISION PROCESSES}\label{mdps-sect}


We begin with Markov decision processes (MDPs)~\cite{Put94},
which are a classic model for decision making under uncertainty.
This is a discrete-time model, with discrete sets of states and actions, that allows both \emph{non-determinism},
e.g., to represent the choices made by the controller of a robot or vehicle,
and discrete \emph{probabilistic} choice, to model environmental uncertainty arising due to,
for instance, the presence of humans, noisy sensors, unreliable communication media or faulty hardware.

We give a formal definition of MDPs below.
Here, and in the remainder of the paper, $\Dist(X)$ denotes the set of (discrete) \emph{probability distributions}
over a finite set $X$, i.e., functions $\mu : X \ra [0, 1]$ such that $\sum_{x \in X} \mu(x) = 1$.
%
\begin{definition}[Markov decision process]\label{mdp-def}
A \emph{Markov decision process} (MDP) is a tuple $\mdp=\mdptuple$ where:
\begin{itemize}
\item
$S$ is a finite set of states and $\sinit\in S$ is an initial state;
\item
$A$ is a finite set of \emph{actions};
\item
$\delta : (S {\times} A) \rightarrow \Dist(S)$ is a (partial) probabilistic transition function,
mapping state-action pairs to probability distributions over $S$;
\item
$\AP$ is a set of atomic propositions and $\lab:S\ra 2^\AP$ is a state labelling function.
\end{itemize}
\end{definition}
\noindent
The execution of an MDP $\mdp$ proceeds as follows. When in a state $s$, there is a non-deterministic choice over the actions that are \emph{available} in the state, defined as  the actions $a \in A$ such that $\delta(s,a)$ is defined and denoted $A(s)$.  
It is assumed that the set of available actions is non-empty for every state. After an action $a \in A(s)$ has been chosen in $s$, 
it is performed and the probability of transitioning to state $s' \in S$ equals $\delta(s,a)(s')$.

\begin{figure}[t]
\begin{minipage}{0.45\linewidth}
\centering
\includegraphics[scale=0.4]{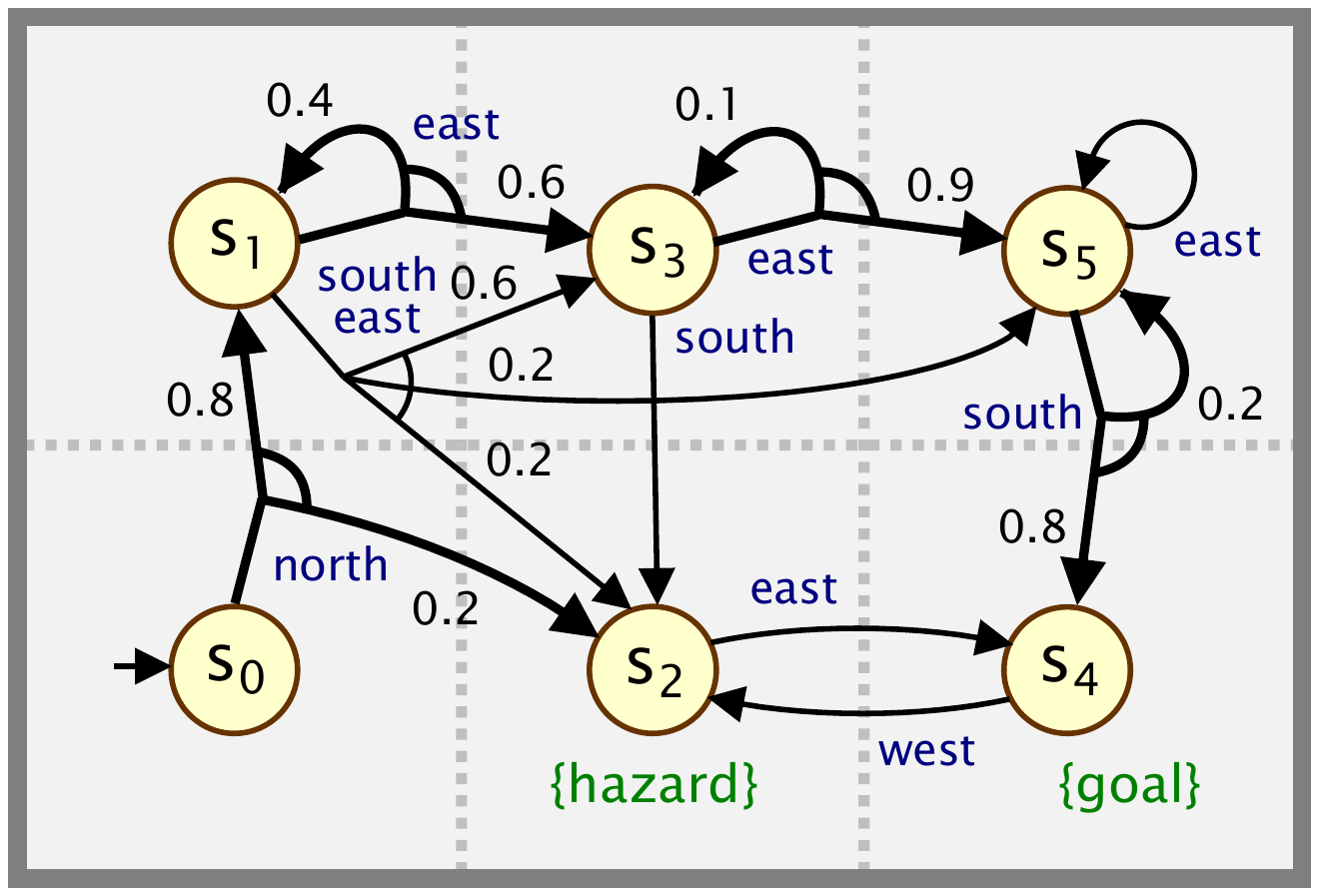}
\end{minipage}
\hfill
\begin{minipage}{0.5\linewidth}
\centering
\includegraphics[scale=0.3]{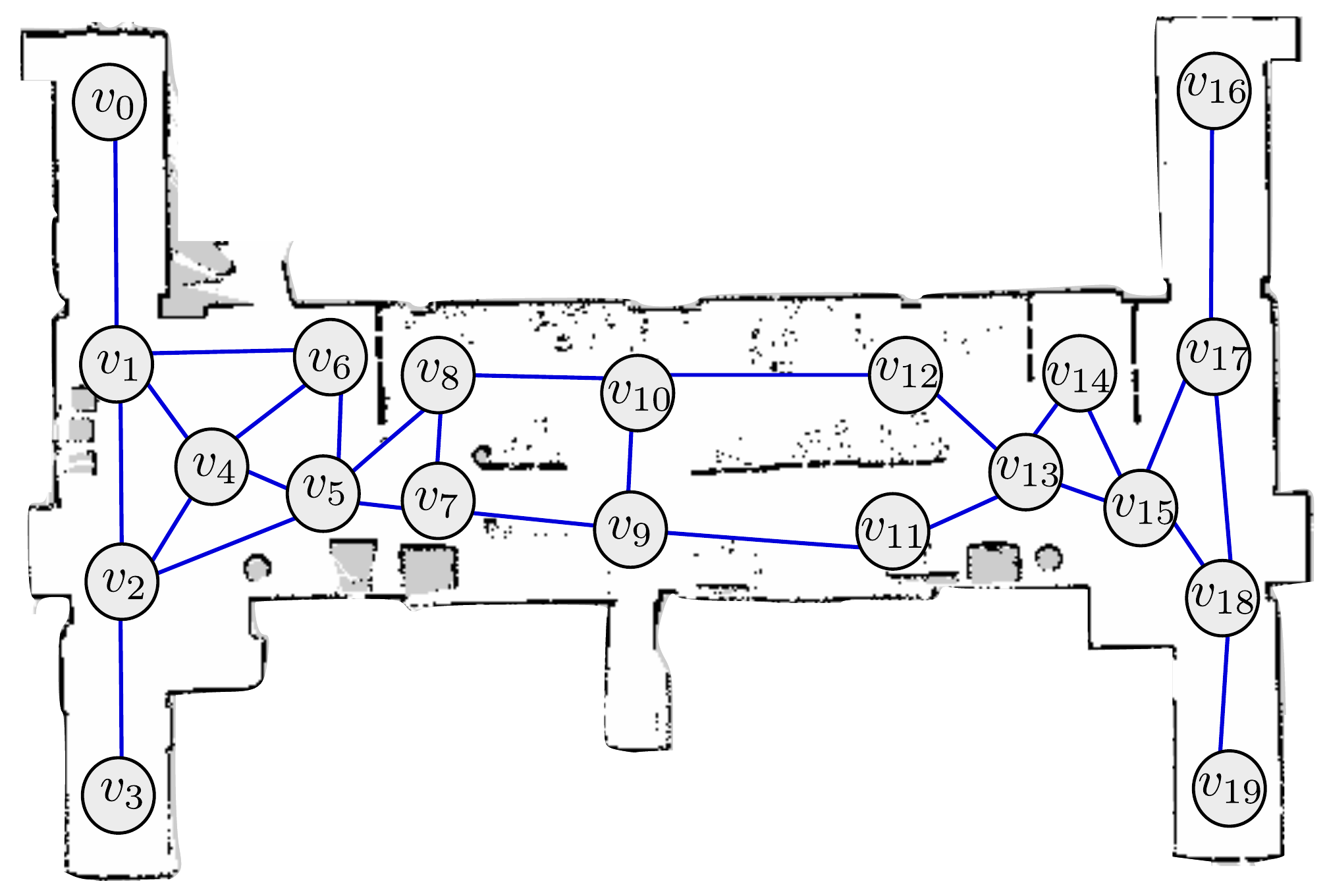}
\end{minipage}
\caption{Left:~A simple MDP representing a robot navigating through a grid; a (deterministic, memoryless) optimal policy for the property $\probop{\max=?}{\neg\mathsf{hazard}\until\mathsf{goal}}$ is marked in bold. Right:~A topological map from \cite{LPH14b} used to build a similar style MDP modelling a mobile robot exploring a building.}\label{mdp-fig}
\vspace*{-0.2cm}
\end{figure}

\begin{example}\label{mdp1-eg}
A simple example of an MDP is shown in \figref{mdp-fig} (left); it models the movement of a robot through locations in a $3\times 2$ grid. Each state ($s_i$) represents a location and actions taken in states result in probabilistic transitions to other locations. For example, in state $s_1$ there is a choice between moving $\mathit{east}$ and $\mathit{southeast}$; if $\mathit{east}$ is chosen, then with probability 0.6 the robot moves east and with probability $0.4$ the robot remains in its current location. Also shown are atomic propositions ($\mathsf{goal}$ and $\mathsf{hazard}$) needed for property specification.
\figref{mdp-fig} (right) shows a topological map used to build a larger, similar-style MDP modelling a mobile robot traversing locations within an office building~\cite{LPH14b}.
\end{example}
\noindent
A path of $\mdp$ is defined by an alternating sequence of action choices and transitions. More formally, a path is a finite or infinite sequence $\pi = s_0 \xrightarrow{a_{\scale{.75}{0}}} s_1 \xrightarrow{a_{\scale{.75}{1}}} s_2 \xrightarrow{a_{\scale{.75}{2}}} \cdots$ such that $s_0=\sinit$, $a_i \in A(s_i)$ and $\delta(s_i,a_i)(s_{i+1})>0$ for all $i \geq 0$. 
$\fpaths_\mdp$ and $\ipaths_\mdp$ denote the sets of finite and infinite paths of $\mdp$, respectively.

We next introduce the notion of a \emph{strategy} (often also called a \emph{policy}) of an MDP~$\mdp$, which
resolves the non-determinism present in $\mdp$.
In particular, strategies decide which actions to take in states of the MDP,
depending on its execution to date.
 
\begin{definition}[MDP strategy]
A \emph{strategy} of an MDP $\mdp$ is a function $\strat:\fpaths_\mdp \rightarrow \dist(A)$
such that, if $\strat(\fpat)(a)>0$, then $a\in\enab{\last(\fpat)}$ where $\last(\pi)$ is the final state of $\pi$.
\end{definition}
\noindent
The set of all strategies of $\mdp$ is denoted $\strats_\mdp$.
We classify a strategy $\sigma\in\strats_\mdp$ in terms of its use of \emph{randomisation} and \emph{memory}.
\begin{itemize}
\item{\bf Randomisation}: $\strat$ is \emph{deterministic} (or \emph{pure}) if $\strat(\fpat)$ picks a single action with probability 1 
for all finite paths $\fpat$, and \emph{randomised} otherwise.
\vskip3pt
\item{\bf Memory}: $\strat$ is \emph{memoryless} if $\strat(\fpat)$ depends only on $\last(\fpat)$
for all finite paths $\fpat$, and \emph{finite-memory} if there are finitely many \emph{modes}
such that, for any $\fpat$, $\strat(\fpat)$ depends only on
$\last(\fpat)$ and the current mode, which is updated each time an action is performed;
otherwise, it is \emph{infinite-memory}.
\end{itemize}
\noindent
Under a particular strategy, the behaviour of MDP $\mdp$ is fully probabilistic and we can reason about the probability of different events.
For a strategy $\sigma$ of $\mdp$, we denote by $\fpaths^\sigma_{\mdp}$ and $\ipaths^\sigma_{\mdp}$ the set of finite and infinite paths that correspond to the choices of $\sigma$. Following~\cite{KSK76}, we can definite a probability measure $\Prob^\sigma_{\mdp}$ over $\ipaths^\sigma_{\mdp}$ that corresponds to the behaviour of the MDP under $\sigma$. Using this probability measure we can then also define, for a random variable $X : \ipaths_\mdp \rightarrow \Rset$, the expected value $\Eset^\sigma_{\mdp}(X)$ of $X$ under $\sigma$.

Random variables can be used to introduce a variety of quantitative properties of MDPs. This is often achieved by augmenting an MDP with reward structures (these can in some cases represent costs, but for consistency we will use the term rewards). Example applications of rewards include: the energy consumption of a device, the number of tasks completed by a robot or the number of packets lost by a communication protocol. 
\begin{definition}[MDP reward structure]\label{mdprew-def}
A \emph{reward structure} for an MDP $\mdp$ is
a tuple $\rew = (\srew,\arew)$, where $\srew : S \ra \Rsetgeq$ is a \emph{state reward function} and $\arew : (S{\times}A) \ra \Rsetgeq$ is an \emph{action reward function}.
\end{definition}

\subsection{Property Specifications for MDPs}\label{mdp-logic-sect}

In order to formally specify the required behaviour of a system modelled as an MDP, we use quantitative extensions of \emph{temporal logic}.
Below, we show a fragment of the logic used as the property specification language for the PRISM model checker~\cite{KNP11}, which we refer to here as \emph{the PRISM logic}. This 
is based on the logics PCTL (probabilistic computation tree logic)~\cite{HJ94} and LTL (linear temporal logic)~\cite{Pnu81}, and also incorporates operators to specify expected reward properties~\cite{FKNP11}.
\begin{definition}[Property syntax]\label{pctl-def}
The syntax for a core fragment of the PRISM logic is:
\[
\begin{array}{rcl}
\Phi &  \; \coloneqq \;  & \probopbp{\psi} \; \mid \; \rewopbr{\rew}{\rho} \\
\psi &  \; \coloneqq \;  & \phi \; \mid \; \neg\psi \; \mid \; \psi \wedge \psi 
\; \mid \; \next \psi \; \mid \; \psi \buntil \psi \; \mid \; \psi \until \psi \\
\rho &  \; \coloneqq \;  & \instant{= k} \; \mid \; \cumul{\le k} \; \mid \; \reachrew{\phi} \\
\phi &  \; \coloneqq \;  & \true \; \mid \; \ap \; \mid \; \neg\phi \; \mid \; \phi\wedge\phi
\end{array}
\]
where $\ap \in \AP$ is an atomic proposition, $\bowtie\,\in\!\{<,\leq,\geq,>\}$, $p \in [0,1]$, $\rew$ is a reward structure, $q \in \Rsetgeq$ and $k\in\Nset$.
\end{definition}
\noindent
Above, we assume that a property $\Phi$ for an MDP comprises a single probabilistic ($\probopP$) or reward ($\rewopR$) operator.
The syntax also includes path ($\psi$) and reward ($\rho$) formulae, both evaluated over paths,
and propositional logic ($\phi$) formulae, evaluated over states.
The intuitive meaning of the $\probopP$ and $\rewopR$ operators, from the initial state of an MDP, is:
\begin{itemize}
\item
$\probopbp{\psi}$ -- the probability of a path satisfying path formula $\psi$ satisfies the bound $\bowtie p$;
\item
$\rewopbr{\rew}{\rho}$ -- the expected value of reward formula $\rho$, under reward structure $\rew$, satisfies the bound $\bowtie q$.
\end{itemize}
A propositional formula $\phi$ is \emph{satisfied} (or \emph{holds}) in a state $s$
if it evaluates to true in that state, where an atomic proposition $\ap$ is true if $s$ is labelled with $\ap$ (i.e., $\ap\in\lab(s)$)
and the logical connectives ($\neg$, $\wedge$) are interpreted in the usual way.

For path formulae $\psi$, the core temporal operators are:
\begin{itemize}
\item
$\next\psi$ (\emph{next}) -- $\psi$ is satisfied in the next state;
\item
$\psi_1\buntil\psi_2$ (\emph{bounded until}) -- $\psi_2$ is satisfied within $k$ steps,
and $\psi_1$ is satisfied until that point;
\item
$\psi_1\until\psi_2$ (\emph{until}) -- $\psi_2$ is eventually satisfied,
and $\psi_1$ is satisfied until then.
\end{itemize}
As is standard in model checking, we use the equivalences
$\future\psi \equiv \true\until\psi$ (\emph{eventually})
and $\globally\psi \equiv \neg\future\neg\psi$ (\emph{always}).
If we restrict the sub-formulae of a path formula to be atomic propositions,
then we get the following common property classes:
\begin{itemize}
\item
$\future\ap$ (\emph{reachability}) -- eventually a stated labelled with $\ap$ is reached;
\item
$\globally\ap$ (\emph{invariance}) -- $\ap$ labels all states;
\item
$\bfuture\ap$ (\emph{step-bounded reachability}) -- $\ap$ labels a state within the first $k$ steps;
\item
$\bgloballyp{k}\ap$ (\emph{step-bounded invariance}) -- $\ap$ labels states for at least the first $k$ steps.
\end{itemize}
Without this restriction, path formulae allow temporal operators to be nested.
In fact the syntax of path formulae given in \defref{pctl-def} is that of linear temporal logic (LTL)~\cite{Pnu81}. LTL can express a range of useful property classes, including:
\begin{itemize}
\item
$\globally\future\psi$ (\emph{recurrence}) -- $\psi$ is satisfied infinitely often;
\item
$\future\globally\psi$ (\emph{persistence}) -- eventually $\psi$ is always satisfied;
\item
$\globally(\psi_1 \ra \next \psi_2)$ -- whenever $\psi_1$ is satisfied,
$\psi_2$ is satisfied in the next state;
\item
$\globally(\psi_1 \ra \future \psi_2)$ -- whenever $\psi_1$ is satisfied,
$\psi_2$ is satisfied in the future.
\end{itemize}
Finally, considering reward formulae $\rho$, the three key operators are:
\begin{itemize}
\item
$\instant{= k}$ (\emph{instantaneous reward}) -- state reward at time step $k$;
\item
$\cumul{\le k}$ (\emph{bounded cumulative reward}) -- reward accumulated over $k$ steps;
\item
$\reachrew{\phi}$ (\emph{reachability reward}) -- reward accumulated until a state satisfying $\phi$ is reached.
\end{itemize}
\noindent
Although omitted from the syntax here for simplicity, it is also common to generalise the third case
and consider the expected reward accumulated until some \emph{co-safe} LTL formula is satisfied.
Intuitively, these are path formulae $\psi$ whose satisfaction occurs within finite time;
examples include ($\future\ap_1)\wedge(\future\ap_2)$ and $\future(\ap_1\wedge\future\ap_2)$,
which require states labelled with $\ap_1$ and $\ap_2$ to be reached,
either in any order (first case) or in a specified order (second case).

\subsection{Probabilistic Model Checking of MDPs}\label{pmc-sec}

Probabilistic model checking is an automated technique for
constructing probabilistic models such as MDPs and then analysing them against
behavioural specifications expressed in temporal logic.
It can be used either to \emph{verify} that a specification is always satisfied, regardless of any adversarial behaviour,
or to \emph{synthesise} a strategy under whose control the system's behaviour can be guaranteed to satisfy a specification.

These ideas are formalised below for the PRISM logic. We first require the following notation.
Satisfaction of a path formula $\psi$ can be represented by a random variable $X^\psi : \ipaths_\mdp \rightarrow \Rset$ where $X^\psi(\pi)=1$ if path $\pi$ satisfies $\psi$ and 0 otherwise.
For a reward structure $\rew$ and formula $\rho$, the random variable $X^{\rew,\rho} : \ipaths_\mdp \rightarrow \Rset$ is such that $X^{\rew,\rho}(\pi)$ equals the state reward or accumulated reward corresponding to $\rew$ and $\rho$ for path $\pi$. 

Verifying that an MDP $\mdp$ satisfies a formula $\Phi$, denoted $\mdp\sat\Phi$,
is defined as follows.

\begin{definition}[Verification problem for MDPs]\label{mdpverif-def}
The \emph{verification} problem is: given an MDP $\mdp$ and a formula $\Phi$, verify whether $\mdp \sat \Phi$, defined as:
\[ \begin{array}{lcl}
\mdp \sat \probopbp{\psi} & \;\;\Leftrightarrow & \;\; \forall \strat \in \strats_\mdp . \, \big( \, \estrat{\mdp}{\strat} (X^\psi) \bowtie p \, \big) \\
\mdp  \sat \rewopbr{\rew}{\rho} & \;\; \Leftrightarrow & \;\; \forall \strat \in \strats_\mdp . \, \big( \, \estrat{\mdp}{\strat}(X^{\rew,\rho})\bowtie q \, \big) \, .
\end{array} \]
In practice, we often solve a \emph{numerical} verification problem:
given an MDP $\mdp$, formula $\probop{\opt=?}{\psi}$ or $\rewop{\rew}{\opt=?}{\rho}$,  where $\opt \in \{ \min,\max\}$,
compute $\estrat{\mdp}{\opt}(X)$ where $X = X^\psi$ or $X=X^{\rew,\rho}$, respectively, and:
\[
\estrat{\mdp}{\min}(X) \rmdef \inf\nolimits_{\strat\in\strats_\mdp} \estrat{\mdp}{\strat}(X)
\quad \mbox{and} \quad
\estrat{\mdp}{\max}(X) \rmdef \sup\nolimits_{\strat\in\strats_\mdp} \estrat{\mdp}{\strat}(X) \, .
\]
\end{definition}
\noindent
Closely related is the strategy synthesis problem.

\begin{definition}[Strategy synthesis problem for MDPs]\label{mdpss-def}
The \emph{strategy synthesis} problem is: given an MDP $\mdp$ and formula $\Phi$
of the form $\probopbp{\psi}$ or $\rewopbr{\rew}{\rho}$,
find a strategy $\strat \in \strats_\mdp$ such that
$\Phi$ is satisfied in $\mdp$ under $\strat$,
i.e., such that $\estrat{\mdp}{\strat} (X^\psi) \bowtie p$
or $\estrat{\mdp}{\strat}(X^{\rew,\rho})\bowtie q$, respectively.

The \emph{numerical} strategy synthesis problem is:
given $\mdp$ and a formula of the form $\probop{\opt=?}{\psi}$ or $\rewop{\rew}{\opt=?}{\rho}$,  where $\opt \in \{ \min,\max\}$,
find an \emph{optimal} strategy $\sigma^\star \in\strats_\mdp$ such that $\estrat{\mdp}{\sigma^\star}(X) = \estrat{\mdp}{\opt}(X)$ for $X = X^\psi$ or $X=X^{\rew,\rho}$, respectively.
\end{definition}
\noindent
For general path formulae, optimal strategies are \emph{finite-memory} and \emph{deterministic}. On the other hand,
for some common cases (e.g., the probability or expected accumulated reward to reach a target),
\emph{memoryless deterministic} strategies are sufficient.
\begin{example}\label{mdp2-eg}
Returning to the MDP from \egref{mdp1-eg}, verification-style queries using the PRISM logic include:
\begin{itemize}
\item
$\probop{\geq 0.8}{\futureop^{\leq 10} \mathsf{goal}}$ -- under all possible strategies, the robot reaches its goal location within 10 steps with probability at least 0.8;
\item
$\rewop{\rew_{\scale{.75}{\mathit{hazard}}}}{\leq 1.5}{\cumul{\leq 20}}$ --  for all possible strategies, the expected number of times that the robot enters the hazard location within the first 20 steps is at most $1.5$;
\end{itemize}
and examples of numerical queries include:
\begin{itemize}
\item
$\probop{\max=?}{\neg \mathsf{hazard} \until \mathsf{goal}}$ -- what is the maximum probability that the goal can be reached while avoiding the hazard location?
\item
$\rewop{\rew_{\scale{.75}{\mathit{steps}}}}{\min=?}{\future \mathsf{goal}}$ -- what is the minimum expected number of steps to reach the goal?
\end{itemize}
Above, we use the following reward structures: 
$\rew_\mathit{steps}$, which  assigns 1 to all state-action pairs; and
$\rew_\mathit{hazard}$, which assigns 1 to all states labelled with atomic proposition $\mathsf{hazard}$.
\end{example}

\subsection{Model Checking Algorithms}\label{mdp-mc-sect}

Probabilistic model checking for MDPs requires a combination of graph-based algorithms, automata-based methods and numerical computation.
The main components of the model checking procedure require computing
optimal probabilities for path formulae and optimal expected values for reward formulae.
For the simplest of these cases (e.g., the probability or expected accumulated reward to reach a target),
various standard techniques for MDPs can be used~\cite{Put94}, including: solving a linear programming problem;  policy iteration (which builds a sequence of strategies until an optimal one is reached); and value iteration (which computes increasingly precise approximations to the optimal probability or expected value).
Of these, value iteration is most commonly used in probabilistic model checking tools,
for scalability and performance reasons,
but variants that offer sound guarantees on the accuracy of results have
also been introduced, e.g.,~\cite{HM18,BCC+14},
as well as methods that employ simulation and heuristics, e.g., \cite{BCC+14,KM20}.
%
%
For finite-horizon (i.e., step-bounded) formulas.
computation of the required values involves a finite number of steps of value iteration.

For reward formula, graph-based precomputation is often also needed.
For example, given a reachability reward formula, a graph-based analysis must first be performed to find the states that reach the target with probability 1 under either all or at least one strategy (depending on whether we are interested in the minimum or maximum expected value).

For more complex path formulae, i.e., full LTL, one must first build a deterministic Rabin automaton (DRA) representation of the path formula and then construct a product MDP consisting of the MDP under study and the DRA. Next, through graph analysis, we  identify states of the product MDP for which the probability of satisfaction is 0 or 1, and the \emph{maximal end components} of the product. Informally, an end component is a set of states for which, under at least one strategy, it is possible to remain in forever once entered and a maximal end component has no other end component as a subset.
After this step, numerical computation is performed on the product in the usual way.

The overall complexity for model checking MDPs against the PRISM logic is doubly exponential in the formula and polynomial in the size of the MDP. However, if we restrict the sub-formulae of path formulae to be atomic formulae, then DRAs are not required and the complexity reduces to linear in the formula and polynomial in the size of the MDP. Further details on the techniques needed to analyse MDPs can be found in, e.g., \cite{FKNP11,BK08,dA97a}
and in standard texts on MDPs~\cite{Bel57,Put94}.

\subsection{Extensions, Tools and Applications}\label{mdp-more-sect}

We conclude our discussion of MDPs by surveying extensions to the basic model checking problems, available software and some practical applications.

\subsubsection{Extensions}
One important extension of probabilistic model checking is to \emph{multi-objective model checking}. This concerns verifying the satisfaction of, or synthesising a strategy that satisfies, multiple properties. The first work in this area concerned multi-objective model checking and strategy synthesis of MDPs against conjunctions of probabilistic LTL specifications~\cite{EKVY08}. The approach has since been extended to general Boolean combinations of LTL properties~\cite{EKVY08,FKN+11} and to include reward formulae~\cite{FKN+11,FKP12}. The synthesised strategies for multi-objective queries have two forms of (finite-)memory: the first corresponds to the satisfaction of the individual objectives and the second, when objectives include general path formulae, the progress towards the satisfaction of such objectives. 

\begin{figure}[t]
\begin{minipage}{0.55\linewidth}
\centering
\includegraphics[scale=1.1]{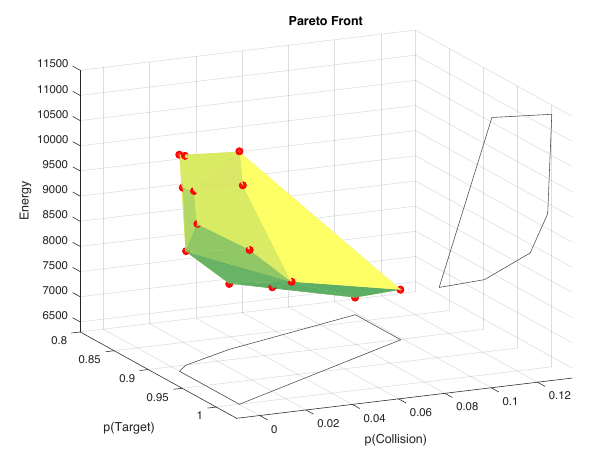}
\end{minipage}
\hfill
\begin{minipage}{0.4\linewidth}
\centering
\includegraphics[scale=1.]{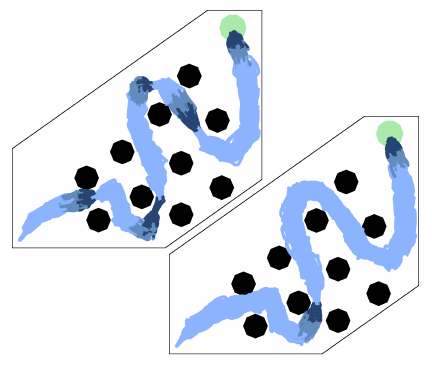}
\end{minipage}
\caption{Left:~Pareto curve, from \cite{LSM18+}, showing the trade-off between three objectives (collision avoidance, target reaching and energy consumption) for different localisation strategies along a particular trajectory of a mobile autonomous robot.
Right: sampled executions for a synthesised strategy.}\label{multi-fig}
\vspace*{-0.2cm}
\end{figure}

Multi-objective model checking has also been extended to numerical queries, which find the optimal value for one numerical objective when restricting to strategies that satisfy the remaining objectives~\cite{EKVY08,FKN+11}. In~\cite{FKP12} this has been generalised to allow the analysis of the trade-offs between objectives by constructing the corresponding \emph{Pareto curve}.
\figref{multi-fig} shows results from \cite{LSM18+}, which uses multi-objective probabilistic model checking of MDPs
to study resource-performance trade-offs in mobile autonomous robots.

Another important extension incorporates parametric techniques. In this approach, one or more aspects of the MDP or specification under study, e.g.\ certain probabilities in the transition function or the step bound in a reward formula, are given as parameters. For Boolean-valued queries, \emph{parameter synthesis} determines the set of parameter values for which the specification is satisfied. For numerical queries, \emph{parametric model checking} returns a symbolic expression for the result, which is a function of the given parameters.

These techniques were first developed for models with only probabilistic behaviour, originally due to~\cite{Daw04} and subsequently extended and optimised in \cite{HHZ11b} and \cite{JCV+14}, which represented transition probabilities as rational functions and applied language-theoretic techniques to return symbolic expressions for reachability probabilities.
This approach has since been  extended to MDPs~\cite{HHZ11} for a subclass of the PRISM logic. An alternative approach applied to MDPs is \emph{parameter lifting}~\cite{QDJ+16}, where parametric transitions are replaced by non-deterministic choices over the extremal values. This non-determinism is placed under the control of a separate player, and therefore the analysis is then performed through probabilistic model checking of a two-player game (see \sectref{tsgs-sect}). 

\emph{Interval MDPs}~\cite{GLD00} generalise MDPs by having interval-valued transition probabilities, and therefore support modelling of systems when there is uncertainty or variation in the probabilistic behaviour. More general notions of such uncertain MDPs allow, for example, \emph{convex uncertainty sets} to represent transition probabilities. Model checking algorithms have been developed for these models on a subset of the PRISM logic~\cite{WTM12,PLSS15}, in a \emph{robust} setting, i.e., where specifications are satisfied for any possible transition probabilities in the allowed set. Extensions to multi-objective queries also exist~\cite{HHH+19}. 

\subsubsection{Tool support}
A number of different software tools are available for model checking MDPs. Probably the most widely used is PRISM~\cite{KNP11}, which supports the logic of \defref{pctl-def} as well as both multi-objective specifications and parametric queries. The tool uses the PRISM modelling language, which is a simple, state-based language, based on Reactive Modules~\cite{AH99}. STORM~\cite{DJKV17} is another tool that supports model checking of MDPs, for a subset of the logic in \defref{pctl-def}, plus multi-objective and parametric extensions, and others such as long run average rewards and conditional probabilities.
Models can be specified in a number of   different modelling formalisms, including the PRISM language and JANI~\cite{BDH+17}. Other general purpose probabilistic model checking tools include the Modest Toolset~\cite{HH14} and ePMC~\cite{HLSTZ14}. PARAM~\cite{HHWZ10} and PROPhESY~\cite{DJN+15} both offer tool support for parametric model checking and synthesis of MDPs.

\subsubsection{Applications}
Applications of MDP-based probabilistic model checking for autonomous systems include: motion planning~\cite{LAB12,LAB15,FGH+20}, spacecraft reconfiguration~\cite{NSTG16}, task allocation and planning for mobile robots~\cite{LPH14b,LFPH19}, analysis of the safety and reliability of robots in extreme environments~\cite{ZRF+19}, human-on-the-loop systems~\cite{LAKG20}, robot battery charge scheduling~\cite{TLHW19} and autonomic computing~\cite{CGKM12}. For a survey on using formal methods (including probabilistic model checking) for the verification of autonomous robotic systems see~\cite{LMD+19}.

%% file: pomdps.tex
\section{PARTIALLY OBSERVABLE MARKOV DECISION PROCESSES}\label{pomdps-sect}

Partially observable MDPs (POMDPs) extend MDPs by restricting the extent to which their current state can be observed,
in particular by the strategies that control them.
In the context of robotics, e.g., it may not be possible to accurately identify a robot's current location
due to either limited precision or unreliability of their sensors.
For security applications, participants in a protocol may rely on the use of private data. 

\begin{definition}[POMDP]\label{pomdp-def}
A POMDP is a tuple $\pomdp = \pomdptuple$ where:
\begin{itemize}
\item
$\mdptuple$ is an MDP (see \defref{mdp-def});
\item
$\obs$ is a finite set of \emph{observations};
\item
$\obsf : S \rightarrow \obs$ is a labelling of states with observations;
\end{itemize}
such that $\enab{s}=\enab{s'}$ for any states $s,s'\in S$ with $\obsf(s)=\obsf(s')$.
\end{definition}
\noindent
In a POMDP, the current state $s$ cannot be directly determined; only the corresponding observation $\obsf(s)\in\obs$ is known.
Notice that \defref{pomdp-def} requires observationally equivalent states to have the same available actions.
This follows from the fact that states that have different sets of actions available would be observationally distinguishable 
as the available actions are not hidden, and hence should not have the same observations.

Above, we adopt a simple notion of observability, used in e.g. \cite{BBG08,CCT13}, which is state-based and deterministic.
More general notions of observations are also commonly used, and may depend on actions performed or are probabilistic.
However, as demonstrated by~\cite{CCGK16}, given a POMDP with these more general notions of observations,
we can construct an equivalent (polynomially larger) POMDP of the form used here.

The notions of paths, strategies, probability measures and reward structures given in \sectref{mdps-sect} for MDPs transfer directly to POMDPs.
The one difference is that the set $\strats_\pomdp$ of all strategies for a POMDP $\pomdp$
only includes \emph{observation-based strategies}. 
\begin{definition}[POMDP strategy]\label{ostrat-def}
A \emph{strategy} of a POMDP $\pomdp = \pomdptuple$ is a function $\strat:\fpaths_\pomdp \ra \dist(\acts)$
such that:
\begin{itemize}
\item
$\sigma$ is a strategy of the MDP $\mdptuple$;
\item
for any paths $\fpat=s_0 \xrightarrow{a_{\scale{0.75}{0}}}s_1 \xrightarrow{a_{\scale{0.75}{1}}}\cdots \xrightarrow{a_{\scale{0.75}{n-1}}} s_n$ and $\fpat^\prime = s_0^\prime \xrightarrow{a_{\scale{0.75}{0}}^{\scale{0.75}{\prime}}} s_1^\prime \xrightarrow{a_{\scale{0.75}{1}}^{\scale{0.75}{\prime}}} \cdots \xrightarrow{a_{\scale{0.75}{n-1}}^{\scale{0.75}{\prime}}} s_n^\prime$
satisfying $\obsf(s_i)=\obsf(s_i^\prime)$ and $a_i=a_i^\prime$ for all $i$, 
we have $\strat(\fpat)=\strat(\fpat')$.
\end{itemize}
\end{definition}

\begin{figure}[t]
\begin{minipage}{0.4\linewidth}
\centering
\includegraphics[scale=0.35]{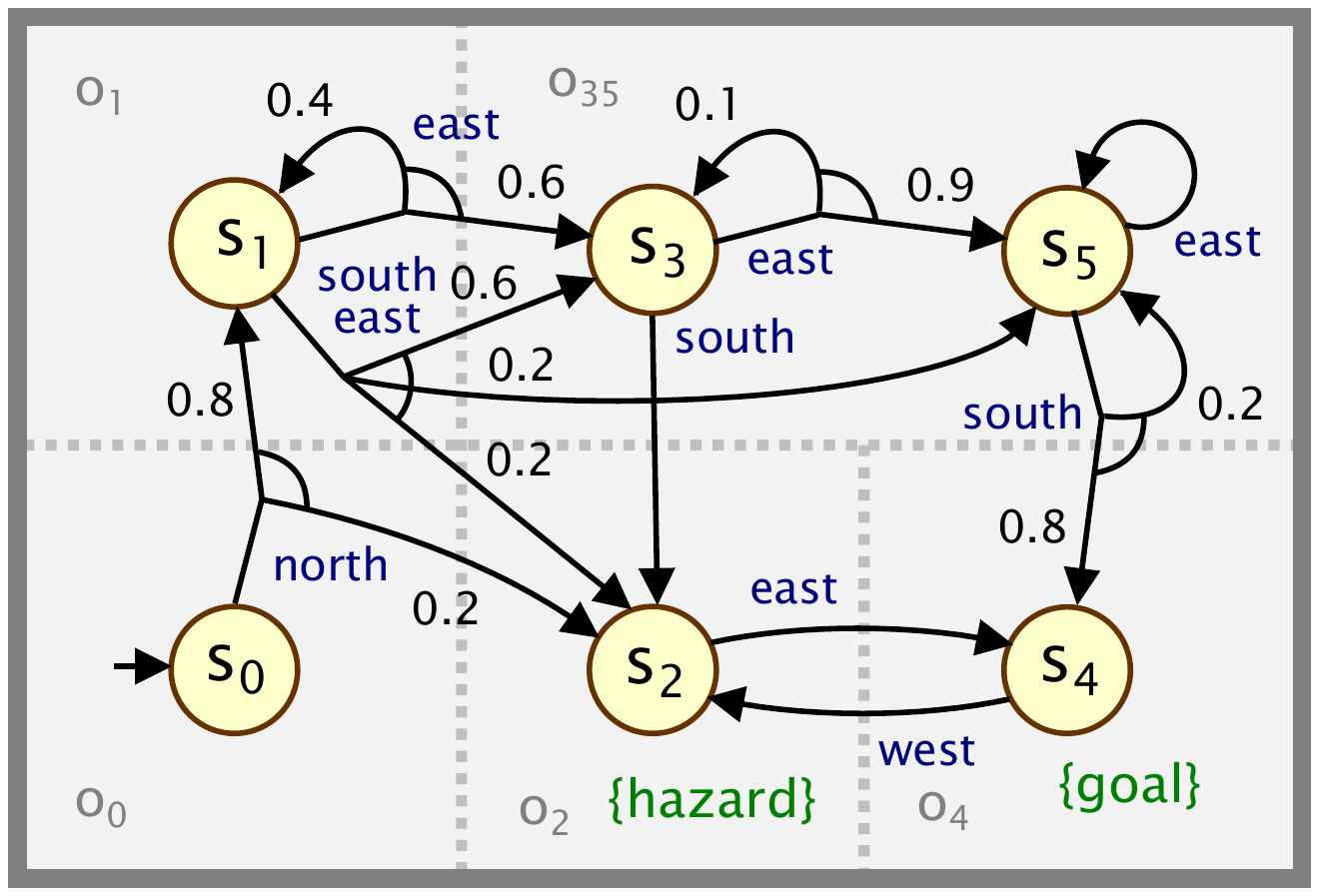}
\end{minipage}
\hfill
\begin{minipage}{0.6\linewidth}
\centering
\includegraphics[scale=0.35]{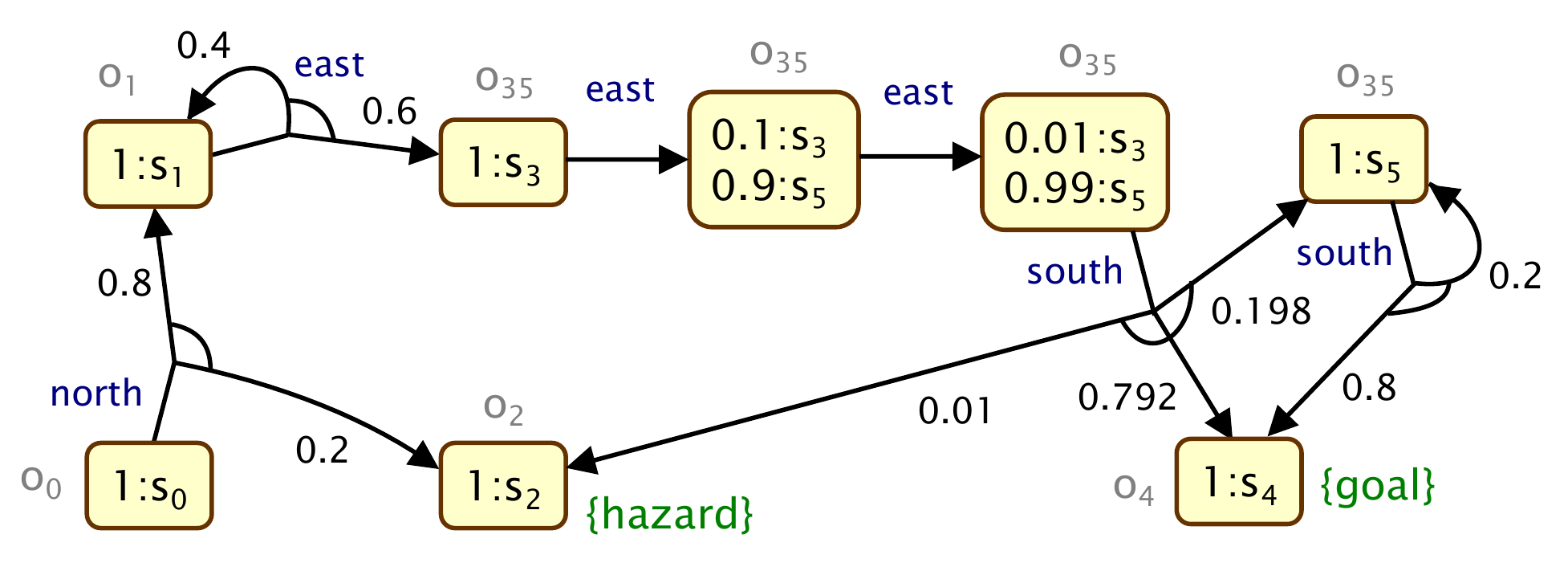}
\end{minipage}
\vspace*{0.5em}
\caption{Left: POMDP variant of the MDP from \figref{mdp-fig}, where states $s_3$ and $s_5$ are observationally equivalent, and therefore cannot be distinguished by strategies. Right: Illustration of the POMDP under the control of a finite-memory strategy; states are labelled with the strategy's current belief as to its current state (as a probability distribution over states).}\label{pomdp-fig}
\vspace*{-0.2cm}
\end{figure}

\begin{example}\label{pomdp1-eg}
\figref{pomdp-fig} (left) shows a POMDP version of the MDP from \egref{mdp1-eg} (\figref{mdp-fig}).
The underlying states, transition probabilities and labels are identical,
but we assume that the grid locations for states $s_3$ and $s_5$ are observationally indistinguishable
due to localisation issues: these states map to the same observation ($o_{35}$), while other states have unique observations ($o_i$ for $s_i$).
\end{example}

\subsection{Model Checking for POMDPs}\label{pomdp-logic-sect}

Properties for POMDPs can be specified using the same logic as for MDPs, presented in \sectref{mdp-logic-sect}. The only change to the verification and strategy synthesis problems (see \defdefref{mdpverif-def}{mdpss-def}) is that the quantification is over observation-based strategies.
However, probabilistic model checking for POMDPs is more challenging than for MDPs since the verification problem for core properties of the PRISM logic is undecidable~\cite{MHC03}.

Verification and strategy synthesis for POMDPs against finite-horizon problems, as well as discounted reward problems, is well studied in the fields of artificial intelligence and planning, and tool support exists, e.g.,~\cite{Pou05}.
However, the PRISM logic incorporates infinite-horizon properties
such as unbounded probabilistic reachability ($\probopbp{\future\ap}$) and
expected reward accumulated to reach a target ($\rewopbr{\rew}{\future\ap}$),
without discounting, where the problem becomes undecidable~\cite{MHC03}.
For further undecidability and complexity results of various POMDP model checking problems, see e.g.,~\cite{BBG08,CCT13}.

Probabilistic model checking of POMDPs was proposed in~\cite{NPZ15}
for a subset of the PRISM logic where path formulae only have propositional formulae as sub-formulae (i.e., without full LTL).
The approach uses grid-based techniques~\cite{Lov91,YB04}, 
which transform the POMDP under study to a fully observable \emph{belief MDP} with uncountably many states
and then approximate its solution based on a finite subset of states (grid points).
Since the problem is undecidable, the approach only returns lower and upper bounds on the quantitative property of interest, and if the bounds are not precise enough, the grid can be refined and analysis repeated.
The efficiency of this approach is improved in~\cite{BJKQ20}
using an abstraction-refinement loop to build smaller MDP approximations.

Strategy synthesis can be incorporated into these methods
based on an analysis of the belief MDP approximation.
The resulting strategies are deterministic but require memory.
Note that these methods assume a fixed initial state (or belief),
in constrast to the methods for MDPs discussed above
which can be performed for all states at once.

Similar methods have been extended to LTL queries~\cite{BTK20}
by translating such formulae to DRAs and using a product MDP construction.
In the qualitative case (checking if the optimal probability equals 0 or 1), under the restriction to finite-memory strategies, model checking algorithms are given in~\cite{CCGK15} for LTL specifications.

Other approaches to POMDP model checking also work by imposing a limit on the
memory available to strategies; this includes~\cite{JJW+18},
which converts the problem to one of to parametric model checking,
and~\cite{WJW+21} which uses a reduction to model checking
for stochastic games using PRISM-games~\cite{KNPS20}.
A related method from~\cite{CJT20} synthesises finite-memory
POMDP strategies represented as recurrent neural networks.

Alternative methods include \cite{CCGK16} which, under the requirement that all rewards in the POMDP are positive, extends approaches developed for finite-horizon objectives to approximate minimum expected reachability rewards.  There is also~\cite{GR12}, which uses counter-example-driven refinement to approximately solve MDPs in which components have partial observability of each other; and~\cite{CCH+11}, which synthesises concurrent program constructs using a search over memoryless strategies in a partially observable stochastic game.

\begin{example}\label{pomdp2-eg}
Consider again the POMDP of \egref{pomdp1-eg} (\figref{pomdp-fig}, left)
and the property specification $\probop{\max=?}{\neg\mathsf{hazard}\until\mathsf{goal}}$.
Any memoryless strategy (i.e., always choosing $\mathit{south}$ or $\mathit{east}$ in both $s_3$ and $s_5$)
has zero probability of achieving this.
\figref{pomdp-fig} (right) illustrates a finite-memory strategy 
for the POMDP of \egref{pomdp1-eg} (\figref{pomdp-fig}, left),
which chooses $\mathit{east}$ twice,
increasing the chance of being in $s_5$, and then $\mathit{south}$.
States are annotated with the current \emph{belief}, i.e., the probability of being in each state.
\end{example}

\subsection{Extensions, Tools and Applications}\label{pomdp-more-sect}

PRISM~\cite{KNP11} implements the algorithms of~\cite{NPZ15} for a subset of the PRISM logic. 
STORM~\cite{DJKV17} also supports POMDP analysis,
via the methods in~\cite{WJW+21,JJW+18,BJKQ20}.
Extensions to synthesise robust strategies for \emph{uncertain} POMDPs,
as discussed earlier for MDPs, can be found in, e.g.,~\cite{SJCT20}.
Applications of POMDP-based model checking for autonomous systems
include robot motion planning~\cite{CCGK15,WJW+21} and human-in-the-loop planning~\cite{CJWT18}. 

%% file: tsgs.tex
\section{TURN-BASED STOCHASTIC GAMES}\label{tsgs-sect}

We now move beyond MDPs to \emph{stochastic games}, which allow for the modelling of cooperative or competitive behaviour
between multiple agents, in the presence of adversarial or uncertain environments.
We start with \emph{turn-based stochastic multi-player games} (TSGs), which have the same structure as MDPs,
except that the states are partitioned amongst a set of players.
Each state is controlled by one player, who resolves the action choices in that state.
Formally, we have the following definition.
 
\begin{definition}[Turn-based stochastic game]\label{tsg-defn}
A  \emph{turn-based stochastic (multi-player) game} (TSG) is a tuple $\tsg = \tsgtuple$, where:
\begin{itemize}
\item
$\mdptuple$ represents an MDP (see \defref{mdp-def});
\item
$N = \{1,\dots,n\}$ is a finite set of \emph{players};
\item
$(S_i)_{i=1}^n$ is a partition of $S$.
\end{itemize}
\end{definition}
\noindent
As for MDPs, in each state $s$ of a TSG $\tsg$, there is a set of available actions denoted $A(s)$, which are the actions $a$ for which $\delta(s,a)$ is defined. However, in this case the choice of which available action is taken in $s$ is under the control of a single player: the unique player $i \leq n$ such that $s \in S_i$. If player $i$ selects action $a \in \act(s)$ in $s$, then, as for MDPs, the probability of transitioning to state $s'$ equals $\delta(s,a)(s')$.

The notion of paths and reward measures are the same as for MDPs. In the case of TSGs we do not have a single strategy, but instead a strategy for each player  $i$ of the TSG that resolves the choice of action in each state under the control of player $i$, based on the game's execution so far. Furthermore, to reason about the behaviour of a TSG we need a strategy for every player, called a \emph{strategy profile}.
\begin{definition}[TSG strategy]
A \emph{strategy} of a TSG $\tsg$ is a function $\sigma_i : \{ \pi \in \fpaths_\tsg \mid \last(\pi) \in S_i \} \ra \dist(\act)$ such that, if $\sigma_i(\pi)(a){>}0$, then $a \in A(\last(\pi))$.
The set of all strategies of player $i \leq n$ is represented by $\Sigma^i_\tsg$ and a \emph{strategy profile} is a tuple $\sigma = ( \sigma_i )_{i=1}^n$ where $\sigma_i \in \Sigma^i_\tsg$ for all $i \leq n$.
\end{definition}
\noindent
Similarly to MDPs, for a TSG $\tsg$ and profile $\sigma$, we denote by $\fpaths^\sigma_{\tsg}$ and $\ipaths^\sigma_{\tsg}$ the set of finite and infinite paths of $\tsg$ that correspond to the choices made by the profile $\sigma$. Furthermore, for a given profile $\sigma$, we can define a probability measure $\Prob^\sigma_{\tsg}$ over the set of infinite paths $\ipaths^\sigma_{\tsg}$ and, for a random variable $X : \ipaths_\tsg \rightarrow \Rset$, we can define the expected value $\Eset^\sigma_{\tsg}(X)$ of $X$ under $\sigma$.

\begin{figure}[t]
\begin{minipage}{0.4\linewidth}
\centering
\includegraphics[scale=0.45]{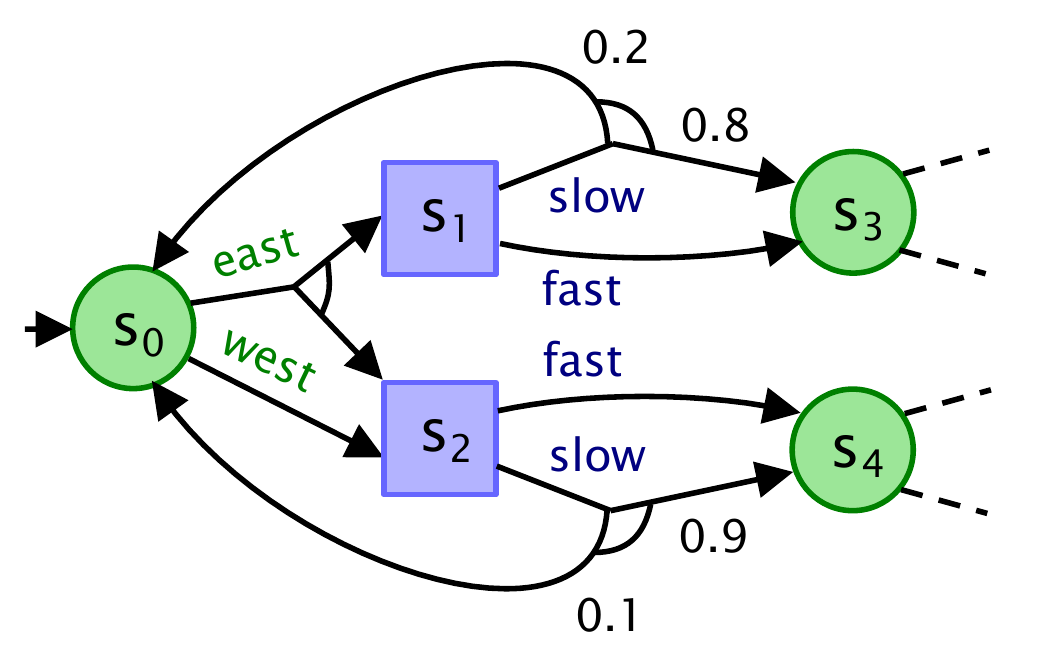}
\end{minipage}
\hfill
\begin{minipage}{0.6\linewidth}
\centering
\includegraphics[scale=0.55]{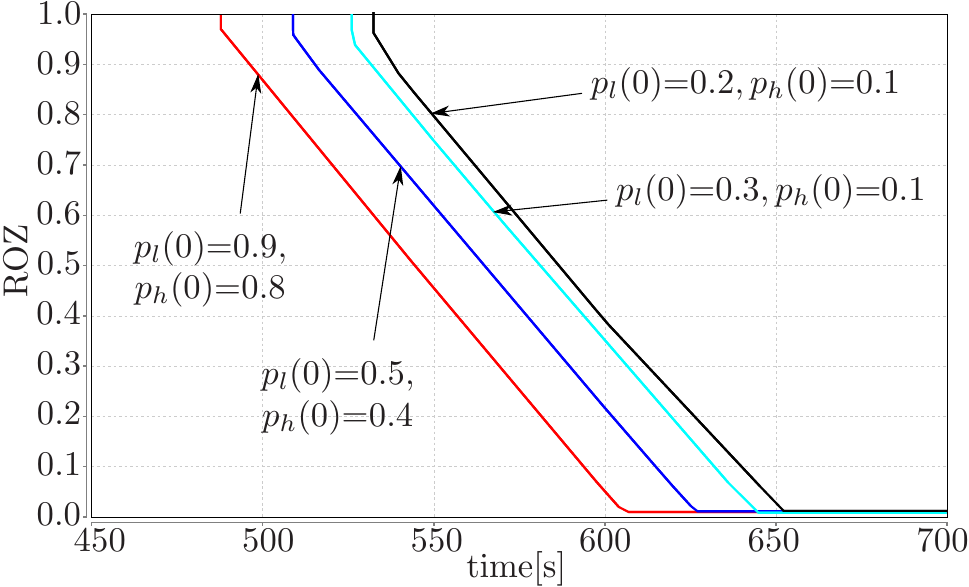}
\end{minipage}
\caption{Left: A simple TSG modelling alternating decisions between a human operator and an autonomous robot.
Right: Results from a more complex, but similar style TSG analysed in~\cite{FWHT16} for an unmanned aerial vehicle partially controlled by a human operator.}\label{tsg-fig}
\vspace*{-0.2cm}
\end{figure}

\begin{example}\label{tsg1-eg}
\figref{tsg-fig} (left) shows a fragment of a simple TSG modelling a human-robot system.
Navigation decisions ($\mathit{east}$ or $\mathit{west}$) are taken by a human operator (circular states, coloured green);
then the robot decides autonomously how to follow these instructions
(square states, coloured blue), here by choosing the speed ($\mathit{slow}$ or $\mathit{fast}$)
with which to proceed.
\figref{tsg-fig} (right) shows results from probabilistic model checking of a more complex TSG model in which an unmanned aerial vehicle performs surveillance under partial control of a human operator~\cite{FWHT16}.
It shows the trade-off between mission time and the likelihood of straying into ``restricted operating zones'' (ROZs)
as operator accuracy varies.
\end{example}

\subsection{Property Specifications for TSGs}\label{tsg-logic-section}

To specify properties of TSGs, we consider an extension of the logic presented earlier for MDPs and POMDPs.
This uses the \emph{coalition} operator $\coal{C}$ from alternating temporal logic (ATL)~\cite{alur2002alternating} to define \emph{zero-sum} formulae.
An extended version of this logic was presented as rPATL (and RPATL*) in \cite{CFK+13b}.%

\begin{definition}[Property syntax for zero-sum games]\label{rpactl-def}
The syntax of extended PRISM logic for zero-sum games is:
\[
\begin{array}{rcl}
\Phi &  \; \coloneqq \;  &
\coal{C} \probopbp{\psi} \; \mid \; \coal{C} \rewopbr{\rew}{\rho}
\end{array}
\]
where path formulae $\psi$ and reward formulae $\rho$
are defined in identical fashion to the PRISM logic in \defref{pctl-def},
$C \subseteq N$ is a coalition of players,
$\bowtie\,\in\!\{<,\leq,\geq,>\}$, $p \in [0,1]$, $\rew$ is a reward structure and $q \in \Rsetgeq$.
\end{definition}
\noindent
The zero-sum formulae $\coal{C} \probopbp{\psi}$ and $\coal{C} \rewopbr{\rew}{\rho}$
can be understood as specifying that the players in the coalition $C$ can collectively ensure that the formula $\probopbp{\psi}$ or $\rewopbr{\rew}{\rho}$, respectively, is satisfied,
against all possible strategies of the players in the set $N \setminus C$.
In order to formalise the semantics of the extended PRISM logic,
for a TSG $\tsg$ and coalition $C$,
we denote by $\tsg_C$
the \emph{coalition game}, that is, the 2-player TSG $\tsg_C$ in which the first player makes all the choices of all players in $C$ and the second all players in $N\setminus C$.

When model checking TSGs, the verification and strategy synthesis problems coincide,
since checking a property $\Phi$ reduces to showing that there exists a strategy for one coalition of players
that satisfies a property for all strategies of another coalition.
%
%
\begin{definition}[Verification and strategy synthesis problems for TSGs]\label{tsgverif-def} 
The \emph{verification problem is}: given a TSG $\tsg$ and formula $\Phi$, verify whether $\tsg \sat \Phi$, defined as:
\[ \begin{array}{rcl}
\tsg \sat \coal{C} \probopbp{\psi} & \;\;\Leftrightarrow & \;\; \exists \strat_1^\star \in \strats_{\tsg_C}^1 . \, \big( \, \forall \strat_2 \in \strats_{\tsg_C}^2 .\,  \estrat{\tsg_C}{\strat_1^\star,\strat_2}(X^\psi) \bowtie p \, \big) \\
\tsg  \sat \coal{C} \rewopbr{\rew}{\rho} & \;\; \Leftrightarrow & \;\;  \exists \strat_1^\star \in \strats_{\tsg_C}^1 . \, \big( \, \forall \strat_2 \in \strats_{\tsg_C}^2 .\,  \estrat{\tsg_C}{\strat_1^\star,\strat_2}(X^{\rew,\rho}) \bowtie q \, \big) 
\end{array} \]
where $\tsg_C$ is the coalition game of $\tsg$ induced by $C$. The \emph{strategy synthesis problem} is to find and return such a strategy $\sigma_1^\star$.

As for MDPs, in practice the \emph{numerical verification problem} is often solved: given a TSG $\tsg$ and formula $\coal{C} \probop{\opt=?}{\psi}$ or $\coal{C}\rewop{\rew}{\opt=?}{\rho}$,  where $\opt \in \{ \min,\max\}$,  compute:
\begin{equation}\label{tsgopt-eqn}
\val_{\tsg_{\scale{0.75}{C^{\opt}}}}(s,X) = \sup\nolimits_{\strat_1 \in\strats_{\tsg_{\scale{0.75}{C}}}^1} \inf\nolimits_{\strat_2 \in\strats_{\tsg_{\scale{0.75}{C}}}^2}  \estrat{\tsg_{\scale{0.75}{C}}}{\strat_1,\strat_2}(X)
\end{equation}
where $C^{\opt} = C$ if $\opt=\max$ and equals $N \setminus C$ otherwise, and $X = X^\psi$ or $X=X^{\rew,\rho}$ respectively. The \emph{numerical strategy synthesis} problem is to return a strategy   $\sigma^\star_1 \in\strats_{\tsg_{\scale{0.75}{C^{\opt}}}}$ such that $\inf\nolimits_{\strat_2 \in\strats_{\tsg_{\scale{0.75}{C}}}^2}  \estrat{\tsg_{\scale{0.75}{C^{\opt}}}}{\sigma^\star}(X) = \val_{\tsg_{\scale{0.75}{C^{\opt}}}}(X)$.
\end{definition}
%
%
\noindent
For general path formulae, optimal strategies are \emph{finite-memory} and \emph{deterministic}, while for infinite-horizon reward formulae and path formulae with only propositional formulae as sub-formulae \emph{memoryless deterministic} optimal strategies exist.

As \defref{tsgverif-def} shows, for verifying TSGs, the main step is computing the value in \eqnref{tsgopt-eqn}. If we consider the coalition game $\tsg_C$ as a \emph{zero-sum game}~\cite{NMK+44}, where the utility function of player 1 is the random variable $X$ and the utility of the second\footnote{In a zero-sum game the utility of the second player is the negation of the first player's utility.} is $-X$, then it follows that  this game is determined~\cite{Mar98}, and therefore the following equation holds:
\begin{equation}\label{tsgopt1-eqn}
\sup\nolimits_{\strat_1 \in\strats_{\tsg_{\scale{0.75}{C}}}^1} \inf\nolimits_{\strat_2 \in\strats_{\tsg_{\scale{0.75}{C}}}^2}  \estrat{\tsg_{\scale{0.75}{C}}}{\strat_1,\strat_2}(X) = \inf\nolimits_{\sigma_2 \in \Sigma_{\tsg_{\scale{0.75}{C}}}^2} \sup\nolimits_{\sigma_1 \in \Sigma_{\tsg_{\scale{0.75}{C}}}^1} \Eset^{\sigma_1,\sigma_2}_{\tsg_{\scale{0.75}{C}}}(X)
\end{equation}
and \eqnref{tsgopt-eqn} is the \emph{value} of this game~\cite{NMK+44}. 
%
%
Furthermore, using Equation~\ref{tsgopt1-eqn} we have the following equivalences:
\[
\coalition{C} \probopbp{\psi} \ \equiv \ \coalition{N {\setminus} C} \probop{\neg (\bowtie p)}{\psi} \;\;\; \mbox{and} \;\;\;
\coalition{C} \rewopbr{\rew}{\rho} \ \equiv \ \coalition{N {\setminus} C} \rewop{\rew}{\neg (\bowtie q)}{\rho} \, .
\]
\begin{example}\label{tsg2-eg}
Returning to the TSG of \egref{tsg1-eg}, PRISM logic queries for this include:
\begin{itemize}
\item
$\coal{\mathit{human}}\probop{\geq 0.6}{\neg\mathsf{crash} \until \mathsf{target}}$ -- a human controller can ensure the robot reaches its target without crashing with probability at least $0.6$, no matter what the robot does;
\item
$\coal{\mathit{rbt}}\rewop{\rew_{\scale{.75}{\mathit{battery}}}}{\max=?}{\instant{=10}}$ -- what is the maximum expected battery level that the robot can ensure after 10 steps, no matter what the choices of the human controller are?
\item
$\coal{\mathit{rbt}}\rewop{\rew_{\scale{.75}{\mathit{steps}}}}{\geq 3.2}{\future \mathsf{target}}$ -- the expected time that the robot requires to reach the target is at least $3.2$, no matter what choices the human controller makes.
\end{itemize}
\end{example}

\subsection{Model Checking Algorithms for TSGs}

Model checking TSGs against the extended PRISM logic can be performed in a similar manner to MDPs (see \sectref{mdp-mc-sect}) using numerical methods, automata and graph-based analysis~\cite{CFK+13b}. For basic properties such as the probability or expected accumulated reward to reach a target,
numerical computation can be performed using several methods including solving a quadratic programming problem, policy iteration and value iteration~\cite{Con93}.
As for MDPs, variants of value iteration that yield error guarantees have also been developed~\cite{KKKW18}.

For the full logic, including LTL, we can translate path formulae to deterministic parity automata (DPAs) and solve a product model which is a stochastic two-player zero-sum parity game~\cite{CH06}. Graph-based algorithms are presented in~\cite{CH12}. Overall, model checking is doubly exponential in the formula and polynomial in the size of the TSG. Similarly to MDP model checking, when the sub-formulae of path formulae are restricted to propositional formulae (i.e., no LTL), then DPAs are not required and parity winning conditions are replaced with reachability objectives and the complexity reduces to $\mathrm{NP} \cap \mathrm{coNP}$.

\subsection{Extensions, Tools and Applications}\label{tsg-more-sect}

Lastly, we discuss extensions, tools and practical applications for model checking of TSGs.

\subsubsection{Extensions}
Multi-objective model checking has also been developed for TSGs, e.g.,~\cite{CKSW13,BKW17} gives algorithms for the synthesis of $\varepsilon$-optimal strategies for TSGs which almost surely satisfy conjunctions of mean payoffs, ratio rewards and Boolean combinations of expected mean-payoffs. On the other hand, \cite{BKTW14} concerns synthesising strategies of TSGs that almost surely surely maintain the averages of a number of long-run average reward specifications remaining above a given multi-dimensional threshold vector.

\subsubsection{Tools and applications}
PRISM-games~\cite{KNPS20} enables the modelling and analysis of TSGs against the extended PRISM logic (see \defref{rpactl-def}) and multi-objective specifications. Applications using TSGs and PRISM-games to model autonomous systems include autonomous urban driving~\cite{CKSW13}, smart grids~\cite{CFK+13b}, human-in-the-loop planning~\cite{FWH+15,JJK+18}, managing collections of autonomic systems~\cite{GCSG15,GGS20} and self-adaption~\cite{CGSP15}. In addition, GIST~\cite{CHJ+10} allows the analysis of $\omega$-regular properties of TSGs and GAVS+~\cite{CKL+11} is a general-purpose tool for algorithmic game solving including TSGs.

%% file: csgs.tex
\section{CONCURRENT STOCHASTIC GAMES}\label{csgs-sect}

In this section, we generalise TSGs to \emph{concurrent stochastic games} (CSGs), in which players choose their actions simultaneously in each state
and without already knowing the actions being taken by other players. 
This can provide a more realistic model of interactive autonomous agents operating concurrently.
\begin{definition}[Concurrent stochastic game] A \emph{concurrent stochastic multi-player game} (CSG) is a tuple
$\csg = \csgtuple$ where:
\begin{itemize}
\item $\mdptuple$ represents an MDP (see \defref{mdp-def});
\item $N=\{1,,\dots,n\}$ is a finite set of players;
\item $A = (A_1\cup\{\bot\}) {\times} \cdots {\times} (A_n\cup\{\bot\})$ where $A_i$ is a finite set of actions available to player $i \leq n$, $A_i \cap A_j = \emptyset$ for all $i \neq j$ and $\bot$ is an idle action disjoint from the set $\cup_{i=1}^n A_i$;
\item $\Delta \colon S \rightarrow 2^{\cup_{i=1}^n A_i}$ is an action assignment function.
\end{itemize}
\end{definition}
\noindent
As previously, we specify available actions
for a CSG, where  for each state $s$ and player $i$ the set of available actions, denoted $A_i(s)$, equals $\Delta(s) \cap A_i$ if this set is non-empty, and $\{ \bot \}$ otherwise. If each player $i \leq n$ selects action $a_i \in \act_i(s)$ in state $s$,  then the probability of transitioning to state $s'$ equals $\delta(s,(a_1,,\dots,a_n))(s')$.

The notions of paths and reward measures are the same as for MDPs. Similarly to TSGs, there is not a single strategy but instead a strategy for each player $i$ of the CSG that resolves the choices of that player.
\begin{definition}[CSG Strategy]
A \emph{strategy} for player $i$ in a CSG $\csg$ is a function of the form $\sigma_i \colon \fpaths_{\csg} \ra \dist(A_i \cup \{ \bot \})$ such that, if $\sigma_i(\pi)(a_i)>0$, then $a_i \in A_i(\last(\pi))$. We denote by $\Sigma^i_\csg$ the set of all strategies for player $i$ and a \emph{strategy profile} is a tuple $\sigma = ( \sigma_i )_{i=1}^n$ where $\sigma_i \in \Sigma^i_\csg$ for all $i \leq n$.
\end{definition}
\noindent
As for the previous models, given a CSG $\csg$ and profile $\sigma$, we denote by $\fpaths^\sigma_{\csg}$ and $\ipaths^\sigma_{\csg}$ the set of finite and infinite paths of $\csg$ that correspond to the choices made by $\sigma$. Furthermore, for a given profile $\sigma$, we can define a probability measure $\Prob^\sigma_{\csg}$ over the set of infinite paths $\ipaths^\sigma_{\csg}$ and,  for a random variable $X : \ipaths_\csg \rightarrow \Rset$, we can define the expected value $\Eset^\sigma_{\csg}(X)$ of $X$ under $\sigma$.

\begin{figure}[t]
\begin{minipage}{0.42\linewidth}
\centering
\includegraphics[scale=0.35]{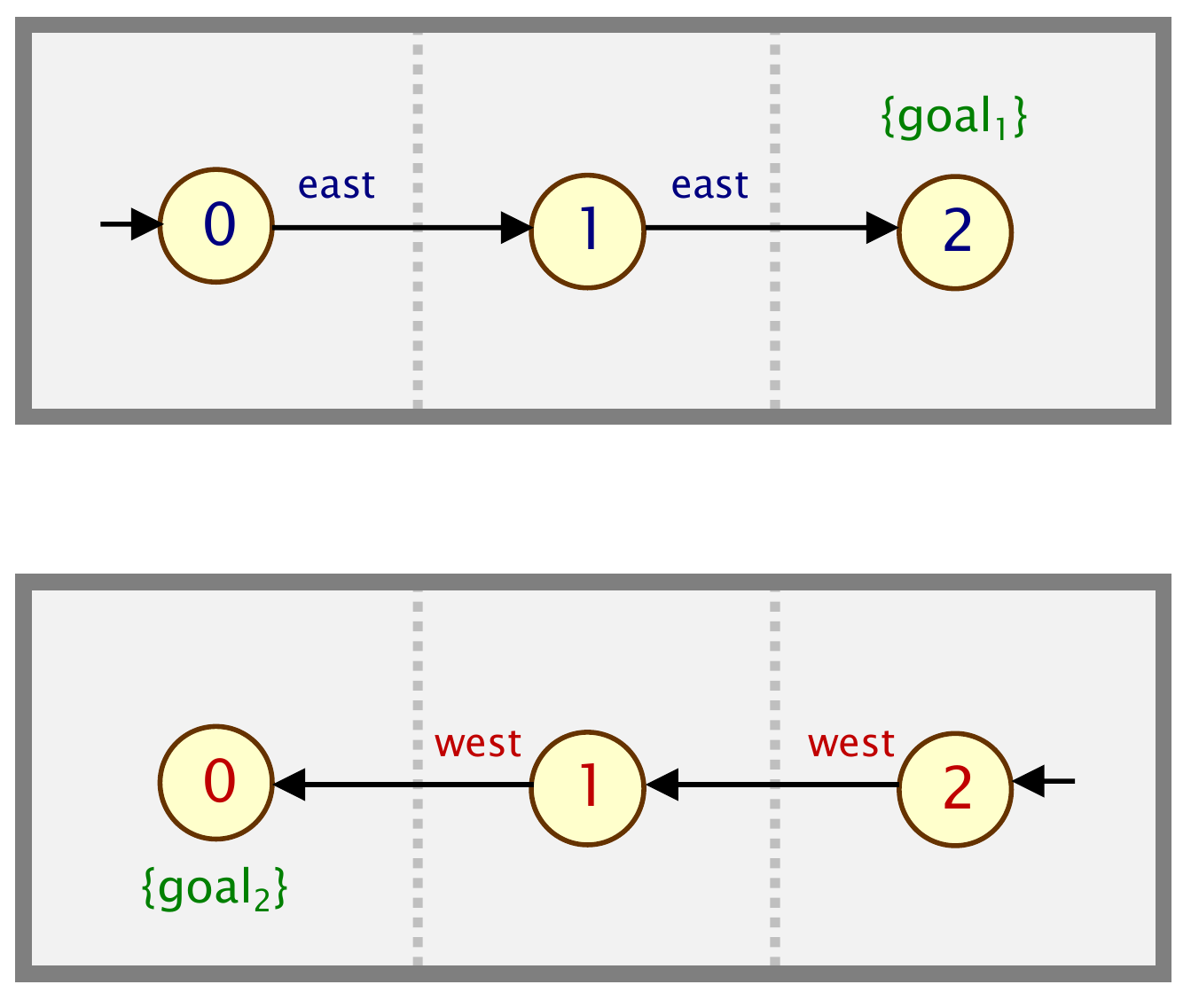}
\end{minipage}
\hfill
\begin{minipage}{0.5\linewidth}
\centering
\includegraphics[scale=0.42]{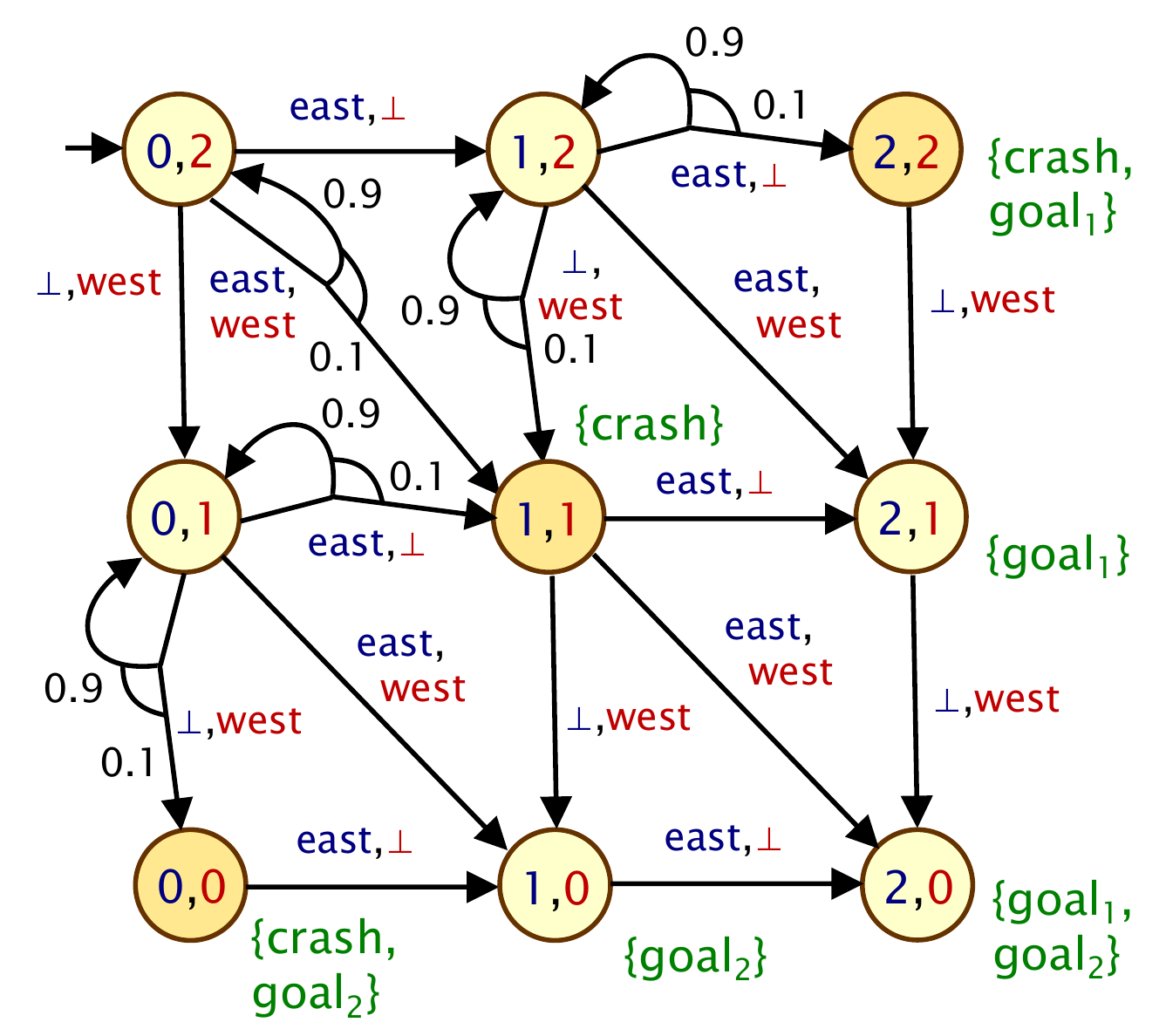}
\end{minipage}
\caption{CSG model of two robots moving through a $3 \times 1$ grid. Left: the transitions of robot 1 (blue) and robot 2 (red).
Right: CSG representing their concurrent execution.}\label{csg-fig}
\vspace*{-0.2cm}
\end{figure}

\begin{example}\label{csg1-eg}
\figref{csg-fig} illustrates a CSG modelling two robots aiming to traverse the same $3\times 1$ grid in opposite directions.
On the left are the transitions of robot 1 (top, blue) and robot 2 (bottom, red).
On the right is the CSG over the product state space (state $(l_1,l_2)$ is when robot $i$ is in location $l_i$), where, if the robots attempt to move to the same grid point with probability $0.1$, they both move and crash into each other\footnote{For simplicity, in this model we assume the robots do not crash when swapping grid points.}, and with probability $0.9$ they do not move. The states of the product are labelled with atomic propositions representing when the robots have reached their goals and if they have crashed.
\end{example}

\subsection{Property Specifications for CSGs}\label{csg-logic-section}

We now extend the logic syntax previously defined for zero-sum games in \sectref{tsg-logic-section}.
We add the specification of \emph{nonzero-sum} properties, using the notion of equilibria,
which allows players to have objectives that are distinct, but not necessarily directly opposing.
\begin{definition}[Property syntax for zero-sum and nonzero-sum games]
The syntax of the extended PRISM logic for zero and nonzero-sum games is:
\[
\begin{array}{rcl}
\Phi & \; \coloneqq \; &  \coalition{C}\probop{\bowtie p}{\psi} \; \mid \; \coalition{C}\rewop{r}{\bowtie q}{\rho} \; \mid \;  \nashop{C_{1}{:}\cdots{:}C_m}{\opt \bowtie q}{\theta} \\
\theta & \; \coloneqq \; &\probop{}{\psi}{+}{\cdots}{+}\probop{}{\psi} \ \mid \  \rewop{r}{}{\rho}{+}{\cdots}{+}\rewop{r}{}{\rho} 
\end{array}
\]
where path formulae $\psi$ and reward formulae $\rho$
are defined in identical fashion to the PRISM logic in \defref{pctl-def}, $C$ and $C_1,,\dots,C_m$ are coalitions of players such that $C_i \cap C_j = \emptyset$ for all $1\leq i \neq j \leq m$ and $\cup_{i=1}^m C_i = N$, $\opt \in \{ \min,\max\}$, $\bowtie\,\in\!\{<,\leq,\geq,>\}$, $p \in [0,1]$, $\rew$ is a reward structure and $q \in \Rsetgeq$.
\end{definition}
\noindent
The logic has been further extended with 
\emph{nonzero-sum formulae} and sums of probabilistic and reward objectives. Nonzero-sum formulae take the form $\nashop{C_{1}{:}\cdots{:}C_m}{\opt \bowtie q}{\theta}$, where
$C_1,\dots,C_m$ are sets of coalitions that represent a partition of players $N$,
and $\theta$ is either the sum $\probop{}{\psi_1}{+}{\cdots}{+}\probop{}{\psi_m}$
of $m$ probabilistic objectives or the sum $\rewop{r_1}{}{\rho_1}{+}{\cdots}{+}\rewop{r_m}{}{\rho_m}$ of $m$ reward objectives. We can consider the $i$th element in these sums as representing the objective $X^\theta_i$ for the coalition $C_i$, where $X^\theta_i = X^{\psi_i}$ or $X^\theta_i = X^{r_i,\rho_i}$, respectively (see \sectref{mdp-logic-sect}).
Their meaning is as follows. The formula $\nashop{C_{1}{:}\cdots{:}C_m}{\max\bowtie q}{\theta}$ is satisfied
if there exists a profile $\sigma^\star$ such that:
\begin{itemize}
\item no coalition $C_i$ for $i \in M$ can deviate from $\sigma^\star$ in order to \emph{increase} their objective $X^\theta_i$;
\item
there is no other such profile for which the sum of the objectives $( X^\theta_i )_{i=1}^m$ is \emph{greater} than the sum under $\sigma^\star$;
\item
the sum of the objectives $( X^\theta_i )_{i=1}^m$ under $\sigma^\star$ satisfies $\bowtie q$; 
\end{itemize}
and we call such a profile that satisfies the first two conditions a \emph{social welfare optimal Nash equilibrium} (SWNE).
The first condition is the standard definition of a Nash equilibrium (NE) profile~\cite{Nas50}
and the second additionally requires that the \emph{sum} of the objectives is maximal, making it an SWNE.
On the other hand, the formula $\nashop{C_{1}{:}\cdots{:}C_m}{\min\bowtie q}{\theta}$ is satisfied
if there exists a profile $\sigma^\star$ such that:
\begin{itemize}
\item no coalition $C_i$ for $i \in M$ can deviate from $\sigma^\star$ in order to \emph{decrease} their objective $X^\theta_i$;
\item
there is no other such profile for which the sum of the objectives $( X^\theta_i )_{i=1}^m$ is \emph{less} than the sum under $\sigma^\star$;
\item
the sum of the objectives $( X^\theta_i )_{i=1}^m$ under $\sigma^\star$ satisfies $\bowtie q$;
\end{itemize}
and we call such a profile that satisfies the first two conditions a \emph{social cost optimal NE} (SCNE). The first condition corresponds to the standard definition of a NE profile for the objectives $(-X^\theta_i )_{i=1}^m$ and the second is what is required for the profile to be social cost optimal. For further details and formal definitions see~\cite{KNPS21}.
To give formal semantics, for a CSG $\csg$ and partition $\cC$ of the players into $m$ coalitions, we denote by $\csg_\cC$ the $m$-player \emph{coalition game}, that is,  the game $\csg$ in which the $i$th player makes the choices for all players in coalition $C_i$.

We can now define the verification and strategy synthesis problems for CSGs which, as for TSGs, coincide as the verification problem reduces to demonstrating the existence of a certain profile. For these problems, we restrict our attention to subgame-perfect NE~\cite{OR04}, which are NE in every state of the corresponding CSG.\footnote{Since the existence of NE is an open problem~\cite{BMS14} for infinite-horizon properties, while $\varepsilon$-NE profiles  have been shown to exist for any $\varepsilon>0$, the definitions are in fact given in the context of a particular $\varepsilon$, see \cite{KNPS21} for details.}

%

\begin{definition}[Verification and strategy synthesis problems for CSGs]\label{csgverif-def}
The \emph{verification problem is}: given a CSG $\csg$ and formula $\Phi$, verify whether $\csg \sat \Phi$, where for zero-sum formulae the satisfaction relation is the same as for TSGs (see \defref{tsgverif-def}) and for nonzero-sum formulae we have:
\[
\begin{array}{rcl}
\csg \sateps \nashop{C_{1}{:}\cdots{:}C_m}{\opt \bowtie q}{\theta} & \;\; \Leftrightarrow & \;\;
\exists \sigma^\star \in \Sigma_{\csg_{\cC}} . \, \big( \, \sum_{i=1}^m \Eset^{\sigma^\star}_{\csg_{\cC}}(X^\theta_i) \, \big) \bowtie q
\end{array}
\]
and $\sigma^\star$
is a subgame-perfect SWNE if $\opt = \max$, and a subgame-perfect SCNE$\,$ if $\opt = \min$, for the objectives $( X^\theta_i )_{i=1}^m$ in the coalition game $\csg_{\cC}$. The \emph{strategy synthesis problem} is then to return such a profile $\sigma^\star$.

The \emph{numerical} verification and strategy synthesis problems for zero-sum formulae are as for TSGs (see \defref{tsgverif-def}): for nonzero-sum formulae, given a CSG $\csg$ and formula $\coalition{C_1,\dots,C_m}_{\opt=?}[\theta]$ where $\opt \in \{ \min,\max\}$, compute, for the objectives $( X^\theta_i )_{i=1}^m$, the sum $\sum_{i=1}^m \Eset^{\sigma^\star}_{\csg_{\cC}}(X^\theta_i)$ for a sub-game perfect SWNE profile $\sigma^\star$ if $\opt = \max$ and for a SCNE profile $\sigma^\star$ otherwise, and then return $\sigma^\star$.
\end{definition}
%
\noindent
Optimal strategies are \emph{finite-memory randomised}, and in both cases \emph{memoryless randomised} strategies are sufficient when restricting to infinite-horizon properties with only propositional formulae as sub-formulae.


\begin{example}\label{csg2-eg}
Returning to the CSG of \egref{csg1-eg}, specifications for this model include: 
\begin{itemize}
\item
$\coal{\mathit{rbt}_1}\probop{\max=?}{\neg\mathsf{crash} \until \mathsf{goal}_\mathsf{1}}$ -- what is the maximum probability with which the first robot can ensure that it reaches its goal without crashing, regardless of the behaviour of the second robot;
\item
$\coal{\mathit{rbt}_2}\rewop{\rew_{\scale{.75}{\mathit{steps}}}}{\leq 4.5}{\future \mathsf{goal}_\mathsf{2}}$ -- there is a strategy for the second robot that can ensure its goal is reached within 4.5 expected steps, no matter the behaviour of the first robot;
\item
$\nashop{\mathit{rbt}_1{:}\mathit{rbt}_2}{\max \geq 2}{\probop{}{\future
\mathsf{goal}_1}{+}\probop{}{\neg \mathsf{crash} \ \untilop^{\leq 10} \mathsf{goal}_2}}$ -- the robots can collaborate so that both reach their goal with probability 1, with the additional condition that the second has to reach its goal within 10 steps and not crash;
\item
$\nashop{\mathit{rbt}_1{:}\mathit{rbt}_2}{\min=?}{\rewop{r_{\scale{.75}{\mathit{steps}}}}{}{\future \mathsf{goal}_1}{+}\rewop{r_{\scale{.75}{\mathit{steps}}}}{}{\future \mathsf{goal}_2}}$ -- what is the sum of expected reachability values when the robots collaborate and each minimises the expected number of steps to reach their goal?
\end{itemize}
\end{example}

\subsection{Model Checking Algorithms for CSGs}

For CSG model checking, only limited progress has been made to date. In the qualitative case, \cite{AHK07,CAH13} present graph-based algorithms for reachability properties and omega-regular languages (which can encode all LTL properties).
In the quantitative case, \cite{KNPS21} introduces model checking algorithms for the extended PRISM logic restricted to a subset of the logic in which the sub-formulae of path formula are propositional formulae and there are only two coalitions in nonzero-sum formulae. There are also restrictions on the class of CSGs that can be analysed, which can be viewed as a variant of \emph{stopping games}~\cite{CFKSW13}.

The model checking algorithms presented in~\cite{KNPS21} involve graph-based analysis followed by value iteration. In the case of zero-sum games, during value iteration for each state, at each iteration, an LP problem of size $|A|$ must be solved (corresponding to finding the value of a zero-sum one-shot game), which has complexity PTIME~\cite{Kar84}. On the other hand, for nonzero-sum formulae, during value iteration for each state, at each iteration, all solutions to an LCP problem of size $|A|$ must be found (corresponding to finding all the NE of a nonzero-sum one-shot two-player game). It has been shown that the complexity of such problems is PPAD (\emph{polynomial parity argument in a directed graph})~\cite{Pap94}. Regarding the number of iterations required in either case, for finite-horizon objectives this is equal to the step bound
in the formula. On the other hand, for infinite-horizon objectives, the number of iterations depends on the convergence criterion and an exponential lower bound has been shown in the worst-case~\cite{HIM11}. 

This approach has since been extended~\cite{KNPS20b} to allow any number of coalitions to appear in nonzero-sum formulae. In this case, during value iteration, for each state, at each iteration, one must find all the NE of a nonzero-sum one-shot $m$-player game, and it has been shown that finding all the NE when there are three (or more) players is PPAD-complete~\cite{DGP09}.

Other work related to CSGs and nonzero-sum properties includes: \cite{CMJ04,Umm10}, which study the existence and complexity of finding NE; \cite{BBG+19}, which analyses the complexity of finding subgame-perfect NE for reachability properties;  and \cite{GNP+19}, which investigates the complexity of equilibrium design. The existence of stochastic equilibria with imprecise deviations and a PSPACE algorithm to compute such equilibria is considered in~\cite{BMS16}.

\vspace*{-0.5em}
\subsection{Tools and Applications}\label{csg-more-sect}

PRISM-games~\cite{KNPS20} supports the model checking of CSGs against a restricted class of the extended PRISM logic,
where the only sub-formulae of path formulae are propositional formulae and nonzero-sum formula are restricted to two coalitions.
An extension of PRISM-games that supports more general nonzero-sum formulae is presented in~\cite{KNPS20b}.
Applications of CSG model checking to date include robotics, computer security and communication protocols such as Aloha~\cite{KNPS21}.

%% file: extensions.tex
\section{FURTHER EXTENSIONS}
\label{Extensions}

In this section, we discuss some further extensions to the models, logics and model checking algorithms to broaden the range of systems and properties that can be analysed.

\subsection{Continuous-Time Models}

The models we have so far presented are all \emph{discrete-time} models exhibiting both probabilistic and non-deterministic behaviour. However, for certain systems it is necessary to also model continuous-time characteristics and the interplay between the
continuous, discrete and stochastic dynamics.

\emph{Probabilistic timed automata} (PTAs)~\cite{Jen96,KNSS02,Bea03} extend MDPs with continuous-time \emph{clocks}, which are variables whose values range over the non-negative reals and increase at the same rate as time. POMDPs, TSG and CSGs have also been extended with continuous-time to POPTAs~\cite{KNPS06}, TPTGs~\cite{KNPS19b} and CPTGs~\cite{FKNT16}, respectively. Reward structures for these models are again specified in terms of both state and action rewards, though state rewards now specify the \emph{rate} at which rewards are accumulated as time passes in a state. 
The PRISM logic can also be  applied to these continuous-time models, where the key difference is that the bounds appearing in path and reward formulae now correspond elapsed time, rather than the number of discrete steps. 
Additional continuous-time properties can be modelled by adding \emph{formula clocks} and \emph{freeze quantifiers} to the logic~\cite{KNSS02}.

Model checking and strategy synthesis algorithms for these models are based on first constructing a finite-state discrete-time model (e.g., an MDP) and then performing model checking on this model. There are a number of approaches that can be employed, including:
\begin{itemize}
\item the region graph construction for PTAs~\cite{KNSS02};
\item the boundary region graph for PTAs~\cite{JKNT09} and CPTGs~\cite{FKNT16};
\item the digital clocks method for PTAs~\cite{KNPS06}, POPTAs~\cite{KNPS06} and TPTGs~\cite{KNPS19b};
\item forwards reachability for PTAs~\cite{KNSS02};
\item backwards reachability for PTAS~\cite{KNSW07,JKNQ17};
\item abstraction refinement with stochastic games for PTAs~\cite{KNP09c}.
\end{itemize}
 
Although this survey focuses on discrete-state models, we also mention briefly that probabilistic model checking
techniques have been developed for models with \emph{continuous} state,
and for \emph{hybrid} stochastic systems with both discrete and continuous aspects to their state space; see, for example, ~\cite{TA13,HSA18}.

\subsection{Verification of Complex Systems}

One of the most successful approaches to improve the scalability of non-probabilistic model checking of complex systems is through \emph{abstraction-refinement} frameworks~\cite{CGJLV00}. This is based on first constructing a small model that abstracts aspects of the complex system 
that do not relate to the specification, while preserving the satisfaction of a given specification. This abstraction can then be verified and, if the abstraction indeed satisfies the specification, 
then so does the complex system. If the abstraction does not satisfy the specification, then information from the model checking process, which in the non-probabilistic case is usually a counter-example path,  can then be used to either show that the specification is not satisfied by the complex system or, if this is not possible, to refine the abstraction until the satisfaction of specification by the complex system can be determined.

In the setting of probabilistic models the focus has been on MDPs. The first framework was introduced by \cite{DJJL01,DJJL02} using probabilistic simulations~\cite{SL95}. Extending this, \cite{WZH07,HWZ08} have developed a framework based on predicate abstraction~\cite{GS97} and probabilistic counter-examples~\cite{HKD09} implemented in the PASS tool~\cite{HHWZ10b}.
An alternative refinement of MDPs is presented in~\cite{KKNP10} using TSGs for the abstract model to maintain a distinction between the non-determinism of the original MDP and that introduced through the abstraction process. By maintaining this distinction, model checking of the abstract models yields separate lower and upper bounds for the given specification, and therefore a quantitative measure of the quality of the abstraction.

An alternative approach to improve scalability is \emph{compositional} verification~\cite{KNPQ13,BKW14}, which allows the correctness of a system to be verified through the model checking of individual components in isolation.

%% file: conclusions.tex
\section{CONCLUSIONS}\label{conc-sect}
We have provided an overview of probabilistic model checking techniques with an emphasis on autonomous system modelling, verification and strategy synthesis from temporal logic specifications. The described techniques have been implemented in the PRISM and PRISM-games probabilistic model checkers and used to model and analyse a variety of case studies from robotics, security and computer networks.   

\begin{summary}[SUMMARY POINTS]
\begin{enumerate}
\item 
Probabilistic model checking provides a unified framework and a formal language to facilitate the construction of a wide range of single- and multi-agent autonomous system models that operate in uncertain or adversarial environments, and whose agents cooperate or compete, and proceed concurrently or in turn-based fashion.
\item 
Temporal logic can be extended with suitable operators (probabilistic, coalition, reward) to conveniently specify a variety of quantitative, zero-sum or distinct, objectives of autonomous agents and supports high-level motion planning,  strategic reasoning and coordination of their behaviours.
\item 
Optimal controllers (or strategies) for autonomous agents (or coalitions of agents), including equilibria, can be automatically synthesised from temporal logic specifications. For partially observable models and infinite-horizon objectives, it cannot be guaranteed that the controllers are optimal due to undecidability of the underlying problem. For multi-agent models, optimality is defined in terms of social welfare and may need to be weakened to $\varepsilon$-optimality for infinite-horizon objectives.
\end{enumerate}
\end{summary}

\begin{issues}[FUTURE ISSUES]
\begin{enumerate}
\item 
Model checking of partially observable stochastic games has been studied~\cite{CD14}. However, little progress has so far been made on developing practical, approximate verification and strategy synthesis algorithms.
\item 
Efficiency is still a major limitation when verifying complex real-world systems. Therefore it is essential to continue to improve scalability, which includes extending compositional probabilistic model checking to all models.
\item 
In the case of CSGs, the efficiency of equilibria computation of normal form games~\cite{Nas50} is the main limitation. Enhancing the performance of this method is thus necessary, as is extending the computation to different notions of equilibria and games including: Stackelberg~\cite{SM73}, correlated~\cite{Aum74} and psychological games~\cite{BD09}.
\item 
One limitation is that none of the presented models allows the modelling of agents that can learn and adapt, e.g.\ through reinforcement learning~\cite{KLM96} or neuro-symbolic reasoning~\cite{dLG09}. This is an important issue that needs to be addressed.
\end{enumerate}
\end{issues}

\section*{ACKNOWLEDGMENTS}
This project has received funding from the European Research Council (ERC)
under the European Union’s Horizon 2020 research and innovation programme
(grant agreement No.~834115) and the EPSRC Programme Grant on Mobile Autonomy (EP/M019918/1).

\section*{DISCLOSURE STATEMENT}
The authors are not aware of any affiliations, memberships, funding, or financial holdings that might be perceived as affecting the objectivity of this review.